**Title: FRAMING CAUSAL QUESTIONS IN LIFE COURSE EPIDEMIOLOGY**

**Authors: Bianca L De Stavola[1], Moritz Herle[2], Andrew Pickles[2]**

**Affiliations:**

[1]UCL Great Ormond Street Institute of Child Health, University College London, London, UK

[2]Department of Biostatistics & Health Informatics, Institute of Psychiatry, Psychology and Neuroscience, King's College London, UK

**Corresponding author:**

Bianca L De Stavola

Population, Policy and Practice Research and Education Department

UCL Great Ormond Street Institute of Child Health

30 Guilford Street

London WC1N 1EH, United Kingdom

**ORCID numbers:**

Bianca de Stavola; 0000-0001-7853-0528

Moritz Herle; 0000-0003-3220-5070

Andrew Pickles; 0000-0003-1283-0346

**Abstract**

We describe the principles of counterfactual thinking in providing more precise definitions of causal effects and some of the implications of this work for the way in which causal questions in life course research are framed and evidence evaluated. Terminology is explained and examples of common life course analyses are discussed that focus on the timing of exposures, the mediation of their effects, observed and unobserved confounders, and measurement error. The examples are illustrated by analyses using singleton and twin cohort data.

# 1. INTRODUCTION

## 1.1. The premise of life course epidemiology

Many acute illnesses and other chronic or recurring impairing conditions that appear in later life are often shaped by developmental processes experienced in utero, childhood, adolescence or early adulthood. For public health policy decisions, knowledge of these processes is often helpful as it may identify the exposures that raise risk or promote resilience. For both early and later ill-health, the aetiological mechanisms are commonly largely hidden from direct observation but must be inferred from analyses of series of observations, typically gathered retrospectively from patients or prospectively from cohorts of healthy individuals. Addressing these research questions is the domain of what is generally referred to as life course epidemiology. In addition to this extended longitudinal perspective, life course epidemiology highlights the importance of a holistic approach to health, as this perspective considers the complex relationships between diseases, not just because of shared risks but because of potentially linked causal chains.

## 1.2. Evidence

There is accumulating evidence that many diseases typically diagnosed in adulthood have social and physiological antecedents much earlier in life. One of the original explanations for this is the Barker hypothesis, which drew a link between foetal nutritional experience and later adult heart disease (Barker and Osmond 1986). In the years since, this has become much generalized into the foetal origins of adult disease (FOAD) (Gillman 2002) and the disease origins of health and disease (DOHaD) paradigms (Bianco-Miotto et al. 2017), both identifying long-term associations that include examples from most areas of medicine. In mental health placing the origins of adult conditions in childhood experience has been seen as important from the earliest days of psychiatry (Breuer and Freud 2004) and though the proposed exposures and mechanisms have changed, the early origins of liability for many mental health conditions remain accepted (Mayer et al. 2021). Despite various forces that press for increasing medical specialisation, we now recognise the full extent of the co-occurrence of multiple chronic conditions and the more limited response to treatment for any one condition in the presence of another (Academy of Medical Sciences 2018).

## 1.3. Implications

While description of patterns and associations over the life course can be fascinating, a focus on identifying the causal mechanisms underlying ill-health is primarily motivated by the desire to intervene: We are wanting to identify opportunities to intervene to improve population health, whether via clinical treatments, behavioural interventions, or policy changes. While reductions or removal of risk factors may be the obvious target, a life course perspective may suggest other opportunities such as displacing exposure to a time when the individual is more resilient or providing compensatory beneficial exposures. Alternatively, it may be possible to interrupt the disease mechanism or reduce the impairments associated with the disease state. These interventions may be directed towards individual patients, groups at high risk, or require changes at the population level, for example by influencing behaviour, such as alcohol drinking habits. The kind of data and designs that will yield evidence of causal effects is likely to differ in each case, and researchers might decide to focus on various more specific research questions, such as testing long-standing hypotheses about the determinants of adult disease, understanding population health trends during rapid demographic transformations, or comparing disease progression with onset of impairment to elucidate causal mechanisms.

## 1.4. Analytical challenges

For any particular problem, the potential solutions to the analytical challenge of attributing cause with sufficient evidential power to justify intervention are many and varied. However, the multidimensional nature of health and well-being in adulthood immediately presents a challenge as to the scope of an investigation and within that, the dimensions to be considered. That disease aetiology may extend over many years, together with the evolving nature of the human organism during the life course, also offers many different ways in which exposure(s) can be characterised, for example cumulatively or pointwise at times that are linked to particular developmental transitions. There may be lags between exposure and evidence of impact, for example incubation or the requirement for a second, or further, exposure (e.g., exposure to stress). Early exposures may increase the risk for later exposures, or an early disease may increase the risk of a second co-occurring one. And even with all these complications, there will likely be not just a single aetiology accounting for all cases, but heterogeneity in both causes that give rise to a common outcome, so-called equifinality, and common patterns of exposures that give rise to variable outcomes, so-called pleiotropy.

These are the challenges presented by the disease process. Numerous additional challenges are invariably presented by shortcomings in the available data. The paucity of early-life confounders, survivor bias affecting studies of disease in later life, measurement error, to name just a few.

The above suggests our task as a daunting one, and sufficiently so such that the methodological failings in most public health research should encourage considerable scepticism. Nonetheless progress is possible. While we focus here on the conceptual clarity and accompanying analytical tools that can help overcome these challenges, we should emphasize that these alone are not enough. Contextual knowledge, strong theory and imaginative use and construction of elegant study designs are as essential.

The paper is structured as follows. In section 2 we review the best-known life course models which are then revisited within a counterfactual-based approach in section 3, where enquiries are refined and estimands introduced. In section 4 we examine the challenges of estimating causal effects when, as it is most likely, assumptions are not met. Here we review possible approaches and also study designs that may be suitable for such tasks. In section 5 we consider the impact of more precise definition of effects on reproducibility and generalisation, key issues for public health policy. We then draw some general conclusions in section 6. Throughout we illustrate issues with examples taken from our applied field, using data form the Avon Longitudinal Study of Parents and Children (ALPAC) (Fraser et al. 2013) and from the Virginia Twin-Family Study of Adolescent Behavioral Development (VTSABD) (Hewitt et al. 1997).

## 2. LIFE COURSE INVESTIGATIONS

### 2.1. Conceptual models

Empirical investigations of early life determinants of later disease or health have traditionally addressed these concerns within a simplified - but broadly accepted- structure. Consider an investigation of the association between childhood socio-economic status (SES) and cognition in later life. Alternative hypotheses for this associations have been proposed (Liu, Jones, and Glymour 2010), in particular apportioning the long lasting effect of childhood SES on cognition to act via its impact on either:

a) education attainment and hence adult occupation, SES, and health behaviours;
b) poor early-life nutrition, and hence poor brain growth and later brain development; or,
c) both.

The first two hypotheses are examples of "*critical period models*" (Liu, Jones, and Glymour 2010; Ben-Shlomo, Mishra, and Kuh 2014) , within which for (a), the critical period is adulthood, and for (b) childhood (Figure 1). However, most empirical investigations of this question have been more consistent with explanation (c): that both periods are relevant. If the latter applies and impact from the two periods is of different intensity, then this hypothesis would be an example of the "*sensitive period model*", with the associations represented by the arrows from either of *A₁ or A₂* into *Y* (A1 → Y or A2 → Y) in Figure 1(c) corresponding to the stronger (more sensitive) association. If instead these associations are of similar magnitude, this would be an example of the "*cumulative exposure model*".

There are still other possible explanations, however. For example, experiencing exposure at later times may enhance the effect of an initial exposure, leading to synergies (or compressions). These might reveal themselves as different final outcomes for individuals with distinct exposure trajectories (e.g., upward, downward, or stable). These would be examples of "*pathway*" or "*chain of risk models*".

### 2.2. Statistical models

There is a strong tradition of comparing these four models formally (Mishra et al. 2009; Smith et al. 2015; Green and Popham 2017), if exposure data at relevant times are available. These formal comparisons rely on their nested nature when viewed in terms of regression models. For example, the most general linear regression model for a continuous outcome *Y* and a single binary exposure *A* observed at two time points on individuals, indexed by *j*, would be,

$$E(Y_j | A_{1j}, A_{2j}) = \beta_0 + \beta_1 A_{1j} + \beta_2 A_{2j} + \beta_3 A_{1j} A_{2j} \qquad (1)$$

where *E(.)* stands for expectation, $\beta_0, \beta_1, \beta_2,$ and $\beta_3$ are unknown regression coefficients. It is easy to see that setting certain constraints on the regression coefficients leads to one of the models described above. For example, setting $\beta_3 = 0$ would imply that there is no synergism between the exposures at different times, and additionally setting $\beta_1 = \beta_2$ would be equivalent to assuming that it is the accumulation of exposure that influences the outcome, irrespective of its timing. If instead both $\beta_2$ and $\beta_3$ were set to be zero, we would assume a critical period model, with $\beta_1$ capturing the effect of early exposure. Fitting model 1 to the available data would allow the formal comparison of these nested models, with the significance (and equivalence) of the relevant regression parameters assessed either via likelihood ratio tests, or other, more flexible, selection criteria.

There are several drawbacks to employing this structured approach to the empirical investigation of life course models. First, the specification of these models should address issues of confounding, especially given the time-varying nature of the exposures. Confounding of an exposure-outcome association occurs when they have (at least) one common cause, adding a non-causal component to their marginal association. Because the processes underlying the evolution of the exposures are most likely to be influenced by other time-varying factors that also influence the outcome of interest, time-varying confounding (also referred to as intermediate confounding) is a particular concern. Ignoring this is likely to lead to substantial biases. Second, it should not be overlooked that most exposures of interest are not just time-varying but also span across multiple, interconnected, dimensions. In the study of cognition described above, adult SES might be understood to capture a broad class of variables, including education, income, and health behaviours, while childhood SES might capture family stability, nutrition, and the home environment. It may also be that this high dimensionality could be conceptualized as evidence of a latent dimension that is the actual driver of the outcome. Ignoring the measurement error in the observed exposure data might therefore also lead to biases. This is especially relevant if the observation times of the exposure do not correspond to the most relevant times in the evolution of the latent drivers (e.g., observed height vs. hormonal spurt experienced in adolescence). Third, the life course models described above may not help address all the questions that one wishes to pose: for example, trying to disentangle whether a critical period or a cumulative exposure model is supported by the data would not help identify the most effective period to implement an intervention (Green and Popham 2017) (see discussion of example 2).

# 3. COUNTERFACTUAL REASONING

## 3.1. Causal diagrams

Recent developments in causal inference offer several tools to deal with these challenges. Causal directed acyclic graphs (DAGs; see Box 1) are particularly useful tools to encode one's assumptions with regards to the processes linking exposures and outcome, including their associations with other (observed and unobserved) variables (Pearl 1995). Once assumptions are encoded in causal graphs that reflect both the nature of the process being investigated and of the available data, empirical investigations of alternative life course models can be initiated. Given the longitudinal nature of the exposures, and the likely considerable number of time-varying confounders involved, such formal graphical representation is very useful for highlighting their complexities. The translation of the conceptual model driving the DAG into relevant targets of estimation, however, is better achieved by adopting the language of counterfactuals, that is of the outcomes that would be experienced under alternative hypothetical interventions on the exposure trajectories.

---

**Box 1**

**Causal directed acyclic graphs (DAGs)**

DAGs are not just a visualization of the conceptual framework (e.g., as shown in Figure 1), but a representation of all possible causal and non-causal relationships involving exposures and outcomes. They are said to be causal when they include all possible common causes of the variables that are included in the diagram, and for this reason are particularly useful to identify sources of confounding bias. Data structures leading to selection bias, such as missing data processes and measurement error for some of the variables, can also be included, as shown in Figures 2(a) and (b) (where *R* is a missing data indicator and $A_2^*$ the unmeasured true exposure at time 2). DAGs can therefore be interrogated to identify the non-causal paths between exposures and outcomes, due to confounding and/or selection bias, and to compare possible strategies to remove ("block") them in order to identify the causal relationships of interest.

---

## 3.2. Potential outcomes and estimands

Let $Y(a_1)$ be the potential outcome, i.e., the value that $Y$ would take if we were hypothetically to intervene on $A_1$ and set it to take the value $a_1$. We can similarly define the potential outcome $Y(a_2)$ for hypothetical interventions on $A_2$. The total causal effect ($TCE$) of $A_j$ on $Y$ can be defined in terms of linear contrasts of mean potential outcomes as

$$TCE_j = E[Y(a_j)] - E[Y(a_j^*)] \qquad (2)$$

with $a_j$ and $a_j^*$ representing the exposed and unexposed status of $A_j$, *j=1,2*, and $E[...]$ indicating expectation over the population of interest.

The $TCE$ is expressed in terms of potential outcomes, that is entities that are not estimable from the data unless certain assumptions are made. Those mostly invoked in causal inference are: no interference and consistency for the exposure, and conditional exchangeability for the exposure-outcome relationship [(Hernan and Robins 2020), Chapters 1 and 3; see Box 2]. Together these assumptions allow us to replace the expectations of potential outcomes that make up the definition of the $TCE$ with functions of the observed data. Estimation can be achieved using for example outcome regression methods (including structural equation models, SEM), propensity score based methods (e.g. inverse probability weighting (IPW) of marginal structural models [(Hernan and Robins 2020), Chapter 11], or double robust methods (e.g. augmented IPW and targeted maximum likelihood estimation; (Hernan and Robins 2020); Chapter 13). Depending on the chosen estimation method, the assumptions of positivity (i.e. that in each stratum defined by the confounders there is a non-zero probability of experiencing either levels of exposure) and correct model specifications (for the outcome or exposure score), are also invoked [(Hernan and Robins 2020) Chapter 11)]. Estimation methods that rely on the availability of instrumental variables replace the NUC assumption with the assumption of homogeneity of the causal effect within the population [(Hernan and Robins 2020) Chapter 16)].

---

**Box 2**

**Identifiability assumptions**

The assumption of no interference is met when the exposure of one individual does not affect the outcome of another; it is thought to be reasonable in most circumstances unless an individual's outcome is affected by interactions with others, as in occurs with vaccination against infectious diseases. Consistency is the notion that clarifies the type of intervention we wish to draw inferences for: it states that the potential outcome for an individual whose exposure is set to take a particular value is the same as the observed outcome for that same individual if they had actually experienced that level of exposure. Assuming consistency implies that we are envisaging interventions that are not invasive (Vanderweele and Vansteelandt 2009).Conditional exchangeability is equivalent to the assumption of no unaccounted confounding (NUC) and states that, within strata defined by the confounders (hence the use of the term "conditional"), the chance of experiencing or not experiencing the exposure is random, i.e. exposed and unexposed individuals are exchangeable (akin to a randomized controlled trials).

---

### 3.3. Causal questions and life course models

Having defined the $TCE$ in terms of potential outcomes, we can now revisit the *critical period model* discussed in the previous section by investigating what would happen to $Y$ if we intervened on $A_1$ while leaving $A_2$ unchanged, in other words what would the effect of $A_1$ that does not involve $A_2$ be: intuitively, its *direct effect*.

Let the potential outcome $Y(a_1, a_2)$ be the value that $Y$ would take if we were hypothetically to intervene on $A_1$ and set it to take the value $a_1$ while setting $A_2$ to take the value $a_2$. Then we could compare $E[Y(a_1, a_2)]$ and $E[Y(a_1^*, a_2^*)]$, with their difference giving what is known as the *controlled direct effect* ($CDE$) of $A_1$ (for a given value $a_2$ of $A_2$), expressed as a linear contrast again:

$$CDE_1(a_2) = E[Y(a_1, a_2)] - E(Y(a_1^*, a_2^*)) \qquad (3)$$

This estimand captures the effect of $A_1$ on $Y$ that does not involve $A_2$ (since we have set the value of $A_2$ to be the same in both expressions). If $CDE_1(a_2)$ were different from zero, for at least some values of $A_2$, there would be evidence for the first period of exposure to be important for the outcome. However, we would not be able to conclude which life course model was supported by the data without further investigation. For example, if $CDE_1(a_2)$ changed for different values of $A_2$, there would be evidence supporting the *pathways model*, as it would be the combination of values of the exposure experienced at different times that would matter.

If $CDE_1(a_2)$ did not vary with $A_2$ (i.e., $CDE_1(a_2) = CDE_1$), and were different from zero, either a *critical/sensitive period* model or a *cumulative exposure model* would be supported by the data. In this case we would want to compare the causal effect of $A_1$ on $Y$ that is not mediated by $A_2$ with the causal effect of $A_2$ on $Y$ (i.e., $TCE_2$, as there are no later, downstream, exposures we are considering here). If $CDE_1 \approx TCE_2$, there would be evidence in favour of a *cumulative exposure* model; if one were larger than the other, the evidence would be for a *sensitive period model*; if only one were (close to) zero, then there would be support for a *critical period model* (Table 1).

Note that, as for $TCE_1$ and $TCE_2$, estimation of $CDE_1$ requires certain assumptions, namely no interference and consistency for the exposure, and NUC for the relationship between $A_1$ and $Y$, and also between $A_2$ and $Y$.

*Example 1*

The following examples serve as illustrations of the different methods described above. It is recognised that the incidence of eating disorders in adolescents is associated with higher body mass index (BMI) in childhood as well as with birth weight (Zehr et al. 2007; Nicholls and Viner 2009; Micali et al. 2018) features that are known to be unequally distributed across different social strata. Thus, to devise preventive strategies, it would be useful to identify whether there are critical or sensitive periods of growth.

To address this question, we use data on female participants from ALSPAC (see Appendix 1). For simplicity, we consider only two periods of exposure, birth and adolescence. Specifically, we use internally standardized birth weight and internally standardized (log transformed) body mass index (BMI) at age 12 years as the exposures of interest ($A_1$ and $A_2$ in model 1), and standardized binge eating score (derived from parental questionnaire data) when the girls were 13.5 year old as the outcome of interest (BE; Figure 3(a)).

Assuming no interference, consistency and NUC (and also correct outcome and mediator model specifications when performing estimation by g-computation), we estimated that the $TCE$ of BMI at 12 on binge eating ($TCE_2$) is to increase 0.24 SD per 1SD increase in BMI (95% confidence interval (CI): 0.21, 0.28), and that the $CDE_s$ of birth weight ($CDE_1(a_2)$), when BMI at 12 is set to take the mean value of 0 or +/- 1 SD are small (in each case it is about 0.02SD change per SD increase in birth weight- around 0.5kg- with small variations). It appears therefore that adolescence is a critical period for the onset of binge eating, while exposure in early life (as captured by birth weight) seems to have a limited impact, although it increases when adolescent BMI is also greater (Table 2).

One should not ignore the possibility that there may be variables on the causal pathway from birth weight to binge eating that act as confounders of the latter's relationship with BMI at 12 (such as $L$ in Figure 3(b)). This source of confounding should be taken into account when estimating $CDE_1(a_2)$,. One such variable could be earlier BMI, as this variable is characterized by strong tracking (Bogin 1999). Re-estimating $CDE_1(a_2)$) while controlling for BMI at 7 leads to slightly increased estimates but does not change our earlier conclusions (Table 2).

## 3.3. Mechanisms

Simply comparing life course models of this sort may leave many questions unanswered. For example, even if we found strong evidence in favour of a critical period model (as per example 1), we would still not be able to say when an intervention might be most effective. This is because the value of the exposure at later times might depend on several earlier factors and intervening on the earlier ones may be more effective than intervening on later ones (Green and Popham 2017).

Specifically, we might wish to quantify what the consequences would be if we intervened on the exposure at selected times after initial exposure and shifted the population distribution of the exposure at those times in a beneficial direction. To do this we consider two new estimands: the *interventional direct and indirect effects* ($IDE$ and $IIE$; (Didelez, Dawid, and Geneletti 2006; Vanderweele, Vansteelandt, and Robins 2014; Vansteelandt and Daniel 2017). These can be viewed as a variation of the *natural direct and indirect effects* (Robins and Greenland 1992; Pearl 2001) that avoids their strongest identifying assumption. This is the "cross-world independence assumption", which implies the absence of intermediate confounding, an assumption that is hard to justify in most life course settings for which multiple causal processes often interact with each other.

The attraction of natural effects is that they allow the partitioning of the $TCE$ into causal pathways that involve and do not involve the mediator. In simple linear settings for both mediator and outcome these estimands are numerically equivalent to the direct and indirect effects most familiar to users of SEM (Daniel and De Stavola 2019), although they lack the formality (and clarity) given by the counterfactual definitions given above.

### 3.3.1. Interventional effects

Assume the causal diagram of Figure 4(a) is correct, with $C$ representing a set of baseline confounders, and $L$ a set of intermediate confounders. We could then imagine a world where the distribution of $A_2$ was changed to resemble that which would occur had $A_1$ been set to take the beneficial value $a_1^\dagger$ (e.g., low deprivation). Let $A_{2|C,L}^\dagger$ be a random draw from that distribution, conditional on confounders $C$ and $L$, and $A_{2|C,L}^*$ a random draw from the distribution of $A_2$, had $A_1$ been set to take the harmful value $a_1^*$ (e.g., high deprivation). Further let $Y(a_1^*, A_{2|C,L}^\dagger)$ be the potential outcome had $A_1$ been set to take the harmful value $a_1^*$ and $A_2$ to take the randomly drawn value $A_{2|C,L}^\dagger$, with an equivalent definition for $Y(a_1^\dagger, A_{2|C,L}^\dagger)$. Then, when expressed as linear contrasts of mean potential outcomes, $IDE$ and $IIE$ are defined as:

$$IDE = E[Y(a_1^*, A_{2|C,L}^\dagger)] - E[Y(a_1^\dagger, A_{2|C,L}^\dagger)] \qquad (4)$$
$$IIE = E[Y(a_1^*, A_{2|C,L}^*)] - E[Y(a_1^*, A_{2|C,L}^\dagger)] \qquad (5)$$

These interventional estimands capture the impact of changing distributions, instead of intervening on individuals (which is what the definitions of natural effects would envisage). For this reason, they are thought to be more useful for assessing the impact of public health policies. Their identification does not require the cross-world independence assumption invoked for natural effects. However, unlike natural effects, their sum does not necessarily equal the $TCE$ of the exposure (and thus cannot be used to quantify mediated proportions).

Interventional direct effects resemble $CDEs$ (compare equations 3 and 4): where they differ is in terms of how the second (intermediate) exposure is set. Because the definition of interventional effects involves a random draw from a chosen distribution, while the definition of $CDE$ involves a preselected value, interventional effects are better suited for continuous exposures (earlier in example 1 we selected three values for BMI at 12 years and hypothetically assigned those same values to everybody which is not realistic at all). However, identifying interventional effects requires additional assumptions to those invoked for the identification of $CDEs$: no interference and consistency for the mediator, and NUC for the exposure-mediator relationship. Estimation can be achieved using a selection of approaches (Daniel and De Stavola 2019).

*Example 1 revisited*

Consider now the question of what the impact on adolescent binge eating scores would be, if we intervened on birth weight as opposed to intervening on BMI at 12. The estimated effect of 1SD increase in birth weight is to increase binge eating scores by about 0.05SD ($TCE$=0.047, 95% confidence interval (CI): 0.016, 0.078; Table 3). This undesirable impact is small but could be much reduced ($IDE$=0.023, 95% CI: 0.010, 0.038; Table 3) if we could intervene on BMI at 12 and change its distribution to replicate that expected had birth weight not increased by 1SD (conditionally on confounders).

These estimates were obtained after specifying flexible models for the outcome and BMI at 12 (with non-linear terms and several interactions) and assuming no interference and consistency for exposure and mediator, and NUC for the exposure-mediator, mediator-outcome and exposure-outcome relationships.

If we had simply assumed linear relationships, and fitted a linear SEM to the same data, we would have obtained much smaller $TCE$ and $IDE$ estimates (Table 3).

### 3.3.2. Multiple mediators

So far, we have considered only two possible exposure periods, and just one dimension of exposure. Life course investigations such as those framed within the DOHAD approach, are more complex than this and involve multiple interlinked and time-varying exposures. Pursuing them requires studying chains of exposures, in other words, dealing with multiple mediators.

In general, unless each mediator acts along distinct pathways, or all relationships are linear, it is not possible to identify the specific contributions of the various mediators. Some interesting questions can however be addressed using a multiple mediator generalization of the interventional effects that does not require a priori knowledge of the causal order among the mediators, nor that there is no unmeasured confounding among them (Vansteelandt and Daniel 2017). This has several advantages because in most situations our substantive knowledge is limited. Being agnostic about their order is possible because these estimands measure the effect of shifting the distribution of each mediator in turn from their counterfactual distribution under no exposure to that under exposure, while the other mediator(s) are set to take values drawn from their marginal distribution under no exposure (and also setting the exposure to exposed status). Thus, if other mediators act downstream from the mediator being studied, they will *not* contribute to its indirect effect.

The interventional direct effect measures the effect of the exposure on the outcome that involves none of the mediators by holding their joint distribution to be that under no exposure. Note that the sum of the interventional effects specific to each mediator is not the same as the interventional indirect effect of all mediators together because the latter involves shifting the joint distribution of all mediators. The difference ("the remainder") is however in most cases minimal and, once accounted for, the sum of the direct and all indirect effects plus this residual term gives the total effect of the exposure.

*Example 2*

In Example 1 we found that adolescence was a critical period for how body size influences binge eating scores. We now consider a range of mediators, as shown in Figure 4, and wish to study how early and later childhood BMI contribute to the association of birth weight with binge eating scores. We first study the joint mediated pathways involving childhood BMI and then focus on separating effects that involve BMI from 7 to 9 (and upstream) and BMI from 10 to 12 (and upstream) using g-computation.

The estimated $IDE$ is 0.018 (95% CI: -0.004, 0.040; Table 4). This represents the extent of the $TCE$ due to an increase of 1SD in birth weight that would remain, if the joint distribution of the five mediators were set to be that of children without that increase. By complement, the estimated $IIE$ involving all mediators is 0.027 (95% CI: 0.017, 0.037) and represents the extent by which the binge eating score would increase if the five mediators were set to have the same joint distribution as that of children whose birth weight was set to be, or not to be, shifted by 1SD, while their birth weight had not changed.

When the distinct pathways involving each of the subgroups of mediators ($M_1$=[BMI7, BMI9] and $M_2$=[BMI10, BMI11, BMI12]) were examined we found that the interventional indirect effect involving BMI from age 10 to 12 was the largest contributor to the indirect effects ($IIE_2$=0.029, 95% CI 0.017, 0.041; Table 4). Because this effect captures the impact of mediators upstream from these ages, and since BMI has such strong tracking, $IIE_2$ includes the impact of earlier BMI. In contrast $IIE_1$ is negligible, indicating no contribution of BMI from 7 to 9 to the outcome that does *not* involve later BMI.

An equivalent partitioning can be achieved by assuming linear relationships among all the variables in the diagram (as one would with a linear SEM). The results are on the right-hand side of Table 4. As observed before, estimates of the $TCE$ and $IDE$ appear to suffer from downward bias because of model misspecification.

### 3.3.4. Other estimands

Sometimes we are not prepared to make the assumptions of NUC for the exposure-mediator(s) relationships. Also, crucially, certain hypothetical interventions may not be well defined. This is the case for birth weight, BMI, and SES (VanderWeele and Hernan 2012; Naimi and Kaufman 2015): for each of these exposures we could construe interventions that would change their value but that would have very different consequences for the outcome, and therefore for the validity of the consistency assumption.

To avoid invoking the consistency assumption for these exposures the mediation question could be rephrased as: "*To what extent the exposure-outcome association can be reduced if we intervened on the mediators?*". An estimand that targets this question is the *counterfactual disparity measure.* It is defined as:

$$CDM(a_2) = E[Y(a_2)|A_1 = 1] - E[Y(a_2)|A_1 = 0] \qquad (6)$$

The $CDM$ is identifiable under the assumptions on NUC for the mediator-outcome relationships and consistency only for the mediator. This definition has been extended to deal with multiple mediators in the form of interventional disparity measures, allowing for the partitioning of the indirect contributions by multiple sets of mediators to the disparity associated with an exposure (Micali et al. 2018).

# 4. POTENTIAL BIASES

The causal quantities (estimands) described in the previous section are useful for addressing some of the most compelling questions posed by life course investigations. However, their identification and estimation require invoking strong and generally unverifiable assumptions such as consistency (of exposures and mediators) and NUC for the relationships involving exposures, mediators, and outcome. In most applications they also invoke the assumption of correct model specifications (although the models requiring correct specification will depend on the estimation method). The importance of the latter was highlighted in the two examples so far, where incorrectly assuming linear models led to smaller (most likely biased) estimates. Incorrect model specifications however can also derive from unaccounted measurement error in the data. As already highlighted, a major source bias would also be introduced by unaccounted confounding. In the next section we focus therefore on the biases that may derive from measurement error and unaccounted confounding, and discuss how one could attempt to address, and hopefully reduce, their impact.

## 4.1. Measurement Error and Misclassification

### 4.1.1. Measurement Error in the Exposure

As VanderWeele et al (2012) noted, the problems posed by confounding in the attribution of causal effects have received considerable attention, but those posed by measurement error are discussed rather less (VanderWeele and Hernan 2012). Measurement error can have both systematic (differential) and random (non-differential) components. While the scope for bias arising from systematic errors in an exposure that are correlated with confounders, outcome, or mediators is obvious, rather less well-known are the malign effects of errors uncorrelated with any of these. Where errors are independent, intuition would suggest some loss of power associated with the extent of measurement error together with some expectation that coefficient estimates for mismeasured exposures would be attenuated towards the null. Indeed, in the case of bivariate exposure-outcome analysis this is generally the case. Although the loss of power remains, the availability of an external reliability coefficient can be used to "disattenuate" this estimate, replacing the original biased estimate $\hat{\beta}$ of the regression coefficient β, by $\frac{\hat{\beta}}{r}$ where r is an intraclass correlation coefficient derived from other studies. Similar attenuation arises with misclassified binary exposures, although the attenuation factor is a function of the specificity and sensitivity of the observed exposure, as well as its prevalence. Such neat corrections however are not transportable to the setting where there are multiple error-prone covariates, as the bias can go in either direction (Keogh et al. 2020).

### 4.1.2. Measurement Error in the Mediator

In simple settings involving a continuous (non-differentially) mismeasured mediator, correctly measured exposure and confounders, and a correctly measured continuous outcome, (and no non-linearities), the expected attenuation of the mediator to outcome coefficient leads to a corresponding shift away from the null in the direct effect, since an underestimate of the mediator-outcome relationship impacts on the estimation of the indirect effect and the sum of (natural) direct and indirect effects equates the total effect between exposure and outcome. With an external knowledge of the reliability (ie r) the replacement of the mediator to outcome coefficient by its disattenuated estimate can be used to correct the estimate of the indirect effect in this simple setting. This can then be subtracted from the total effect of the exposure to obtain an unbiased direct effect. These simple adjustment methods can be used when the outcome is modelled using linear regression for continuous outcomes or logistic regression for binary outcomes (le Cessie et al. 2012) with similar results for binary mediators subject to misclassification (Ogburn and VanderWeele 2012).

These adjusted estimates both assume that the reliability coefficients obtained from other studies are applicable and are also known without error. If additional data on the variable affected by error were available, then a joint multivariate analysis of the outcome and mis-measured variable using additional data from an embedded reliability study provides an approach that allows error propagation, since the contribution to the target estimates arising from uncertain reliability can be considerable. Joint analysis with both embedded and external reliability data can reduce this and allows scope to test the stability of reliability across studies.

Loeys et al (2014) consider the estimation of the controlled direct effect where the mediator is considered as latent, measured only indirectly through indicator variables (Loeys et al. 2014). They identify several circumstances where the routine estimation of the direct effect from a SEM model would not provide unbiased estimates of the controlled direct effect, circumstances not just relating to non-linearity but also to the need for correct specification of the exposure-mediator relationship and its confounders. Their proposed two-step g-estimator nonetheless uses standard SEM tools to estimate the values of the latent mediator, but importantly estimated from the fitting of the whole model and not just a factor model to the indicators (Skrondal and Laake 2001). This work has been extended by Loh et al (2020) to controlled direct effects with time-varying mediators and outcomes, estimated by g-estimation (Loh et al. 2020).

These developments are particularly relevant for life course research where growth curve models have become a popular and parsimonious way of characterizing development over a possibly extended period of time by means of a limited number of individual latent growth intercept and slope parameters, typically treating variation around the growth trajectories implied by these quantities as conditionally independent measurement errors. Linked models of this type allow us to consider a mediational process, such as that of Figure 5. Here the trajectories of the repeated measures of both the mediator $M$ and the outcome $Y$ are assumed determined by latent/random intercepts and slopes. The effect of an exposure variable $X$ on the trajectory of $Y$ is decomposed into its effects directly on the growth of the outcome $Y$ that does/does not involve the growth of the mediator $M$.

Sullivan et al (2021) provide counterfactual based derivations for the natural direct and indirect effects of a change in the level of exposure $X$ on $Y$, which are coincident with those that would be calculated for the same model and change in exposure (say $x$ to $x^*$) using the methods of traditional SEM path calculus (Sullivan et al. 2021). However, as discussed in section 3, natural effects require the identification assumption of no intermediate confounding and this is quite unrealistic in life course settings. They do however clarify via their counterfactual derivation the importance of explicating the identification assumptions that are instead left implicit within the traditional SEM framework. There are also differences in the models that are considered eligible by the SEM and counterfactual approach. For the first, MacKinnon (2008) allows the intercept of the outcome $I_Y$ to affect the slope of the mediator $S_M$ (MacKinnon 2008), an effect that Sullivan et al (2021) argue would break the last assumption of those listed above. Indeed, whether using either counterfactual or SEM path calculus, the apparent influence of slope terms, whose values only become apparent post-baseline, on the values of intercepts makes for a certain discomfort in relation to wanting causes and effects to follow a natural temporal precedence. While resolvable by assuming that both intercept and slope are determined and fixed at baseline (e.g. genetically determined) this may be a conceptually more restrictive model than many users had conceived it to be.

*Example 2 revisited*

To illustrate this approach, we revisit the causal enquiry regarding the role of childhood BMI on binge eating scores by focussing on the growth features underlying the observed BMI measures, as shown in Figure 6, namely latent size and latent velocity. These are derived from the information from all ages from 7 to 12 years, after selecting the best fitting specification for the effect of birth weight and age, controlling for the baseline confounders and including several interactions with both exposure and age. By addressing issues of measurement error in the observed BMI values, and adopting g-computation as a generalization of the approach described by Loh et al (2014), we obtain a marginally larger estimate of the interventional indirect effect involving all mediators (0.028, 95% CI: 0.18, 0.38 vs. the earlier estimate of 0.027; Table 4) and consequently a marginally smaller estimate of the interventional direct effect (0.017, -0.005, 0.039 vs. the earlier estimate of 0.018). Of the two mediators, only size appears to contribute to the interventional indirect effect (0.026, 95% CI: 0.016, 0.036). If the relevant assumptions hold, these results indicate that it is not just BMI in later childhood that matters, what we had concluded before, but the average size that is expressed by all BMI measures starting from age 7. This agrees with the earlier results that found that the indirect effect via the later BMI measures and upstream from them was the main pathway of increased scores.

The estimates obtained by fitting a linear SEM are again smaller (Table 4).

## 4.2. Unaccounted confounding

Unaccounted confounding affects the majority of observational studies. The extent of the bias induced by this depends on our understanding of the processes that lead to the observed exposure distribution, as well as the availability of data on the variables that may aid our ability to control the confounding paths. There are however other possible strategies. In the following we discuss some of them.

### 4.2.1. Experimental designs

Experimental designs such as randomized trials are sometimes possible even for long-term public health studies. Step-wedge designs in which an intervention is introduced sequentially over randomised geographical or service units have been promoted (Medical Research Council 2000). Some interventions can be evaluated by the use of encouragement designs, in which rates of take-up are increased by some stimulus, such as a letter (Wardle et al. 2016), a stimulus so modest that of itself it has no impact on the health outcome under investigation but nonetheless offers sufficient encouragement to achieve a modest increase in uptake. Sent to a random subset of participants, the assignment to the encouragement group can be used as an instrumental variable (IV), that allows estimation of the effect of the intervention $A$ on the health outcome $Y$ (Figure 7(a)) that is unaffected by both measured and unmeasured confounders, therefore gaining an interpretation as total causal effect of the intervention, $TCE$, provided that: *(i)* the instrument is relevant, i.e., instrument and intervention $A$ are associated; *(ii)* the instrument does not share any common causes with $Y$ (the marginal exchangeability assumption); *(iii)* the instrument has no other causal pathways to $Y$ besides via $A$ (the exclusion restriction (ER) assumption); and *(iv)* the causal effect is in some form homogeneous (Hernan and Robins 2006). The first and second condition should be met because of randomization, while the third might not be satisfied for example if those receiving the encouragement modify other behaviours because of knowing that they were receiving the encouragement.

An alternative assumption to the fourth assumption of effect homogeneity is the monotonicity condition, informally stating that the relationship between instrument and intervention is in the same direction for all individuals. This would not yield an estimate of the $TCE$, however. Swanson and Hernan (2018) show that the interpretation of the causal effect estimate under this assumption as a local effect (among those who "comply") is only valid when the instrument itself is causal (Swanson and Hernan 2018). This seriously limits the range of useful instruments. It is not surprising that finding instruments for mediators is even more challenging. However , where a moderator of the effects of the intervention on a potential mediator can be found, this can be used as an instrument for the mediator and the set-up of Figure 7(b) allows for mediation estimates with a causal interpretation, assuming effect homogeneity for both instruments (Emsley, Dunn, and White 2010).

### 4.2.2. Natural experiments

There have been many imaginative suggestions for natural sources of variation that mimic the randomised assignment to encouragement of the previous section. A common variable that has influenced treatment participation has been geographical distance to the service (Baiocchi et al. 2010). Though distance has frequently been used as an instrumental variable, the plausibility of the necessary assumptions has been hard to judge. More recently the availability of genotyped cohorts means that measured genetic variation is becoming available as an indicator of variation that could be plausibly assumed to be conditionally random. One exploitation of these data, now widely known as Mendelian Randomisation (MR), has led to numerous publications, some landmark in seeming to resolve major debates of long-standing causal questions (Lawlor et al. 2008; Zheng et al. 2017). In the case of single genes of large-effect the standard Instrument variable (IV) model of Figure 7(a), is directly applicable, but the rarity of the availability of large effect genes (e.g., alcohol dehydrogenease) has required adaptations to the approach. Cumulating sources of genetic variation into polygenic scores as a single instrumental variable reduces the plausibility of the IV assumptions by increasing the risk of pleiotropy – that at least some of the genomic variants in the risk score would not satisfy the exclusion restriction. Retaining the individual variants as a collection of IVs also increases the problem of weak instruments. However, the realisation that larger genetic effects on the exposure should go along with larger effects on the outcome provided scope for not only reducing weak-instrument bias but also testing for pleiotropy and is an approach now known as MR-Egger regression (Bowden, Davey Smith, and Burgess 2015). The genes-as-instruments approach has been a motivation for resolving issues of reverse causality, for example highlighting that rather than C-reactive protein (CRP) being a cause of inflammatory disease it is now considered more likely that its raised levels are a consequence of that disease (Marott et al. 2010). However, the time-varying nature of CRP renders the exclusion restriction assumption most unlikely as it needs to be satisfied for all life-time exposures. Consider Figure 7(c) where for simplicity only two measurements of CRP are shown (as $A_1$ and $A_2$). If studying the causal effect of CRP experienced during this timeframe, the two red arrows from the instrument to $A_2$ and from $A_1$ to $Y$ should be absent. If we were interested only in exposure at the earlier time point, $A_1$, then the arrow labelled 1 should be absent; if we were interested in later exposure then it would be the arrow labelled 2 that should be absent. Generalizing the argument to lifetime exposure (as often referred to in MR studies) it becomes apparent that the ER is unlikely to be satisfied.

If the outcome too were time-varying as shown in Figure 7(d), other sources of bias may affect our causal investigation. Reverse causality could be an issue if there is feedback from exposure to outcome and vice-versa, with the genetic association with the exposure being distorted by reverse causality (Burgess, Swanson, and Labrecque 2021). It follows that for life course studies where exposures and outcomes are dynamic, causal investigations that rely on genetic instruments require strong assumptions and clear specification of the timeframe for both exposures and outcomes.

Finally, It is important to distinguish between genetics – the between individual variation in the DNA sequence considered as relatively immutable – from epigenetic and gene-expression (transcriptomic) variation with which it is commonly discussed. Epigenetic and genetic expression measures can and often do change over time and they do so in response to the environment, and form the core of much research within the DOHAD paradigm (Bianco-Miotto, Craig et al. 2017). Thus, unlike genetic variation, their variation is subject to the same sources of confounding as other measures of exposure. However, as mediators of genetic differences, they become themselves amenable to the confounder control that could be obtained from genes.

### 4.2.3. Using related individuals

In addition to contemporary methods using observed genetic variants, there is a rich history of analysing data from related individuals in health research, including twin studies. The substantially random segregation of maternal and paternal autosomal genes during fertilization, relatives with varying degrees of shared parentage, and the chance division into the genetically identical embryos of mono-zygotic twins provide scope for sources of structured variation that can be exploited to advantage to provide counterfactuals from which causal estimates can be derived. Figure 8 presents a DAG for twin-pairs that we will examine for the effects of life-events ($X$) on subsequent behaviour ($Y_2$) in the presence of confounding effects of prior behaviour ($Y_1$) and unknown shared confounder $U$.

Various estimators have been proposed for exploiting data of this kind but as Sjolander et al (2012) elaborate some of these provide estimators that can be shown not to be causal, while others do not all give estimates of the same counterfactual comparison or require additional assumptions (Sjölander, Frisell, and Öberg 2012). The standard twin model decomposes variation within and between twin pairs in a single phenotype into additive random components of variation for additive (and dominant) genes, shared and non-shared environment. In a setting of two phenotypes, one considered an exposure and the other an outcome, when considered jointly the twin model allows a decomposition of not just the variation and covariation of each phenotype between twins in a pair, but also of the covariance between the phenotypes, including some components that would ordinarily be considered as major sources of unmeasured confounding. In the linear case it is possible to construct a complex and highly parameterised SEM with all the correlated and uncorrelated components variance that follow the standard genetic and environmental effects decomposition of the classical twin model (McAdams et al. 2020). However, most analyses attempt to condition away many of these components and a commonly proposed model is the between-within or between-within (BW) model (e.g., Carlin et al. 2005) that adjusts for all shared confounders.

$$g[E(Y_j|X_j, \bar{X})] = \beta_0 + \beta_W X_j + \beta_B \bar{X},$$

Where $g$ is a collapsible link function such as identity, and the exposure $X$ is continuous, an assumption of linearity of effects across pairs allows the estimate of $\beta_W$, in the absence of non-shared confounders, to be interpretable as a population causal effect ($TCE$). Where the exposure is binary the estimate can only be considered as applying to the discordant sub-population of twins. Where $g$ is a non-collapsible link e.g., logit, there is the further complication arising from the non-equality of the conditional, or subject-specific, and marginal estimates of the effect coefficients.

The BW model is frequently extended to allow for an observed non-shared confounder covariate by the addition of the single variable $V$ measured for the target twin. However, Sjolander et al (2012) show how conditioning on the twin-mean exposure induces a collider bias in the exposure-confounder association but that the inclusion of the measured covariate for both twins as shown below results in unbiased estimates.

$$g[E(Y_j|X_j, \bar{X}, V_j, V_{j'})] = \beta_0 + \beta_W X_j + \beta_B \bar{X} + \gamma V_j + \gamma' V_{j'},$$

*Example 3*

We illustrate these estimation steps using data from the Virginia Twin Study of Adolescent Behavioural Development sample of 733 MZ twins and 376 DZ twins (see Appendix).Using the same-sex twins from the Virginia Twin Study of Adolescent Behavioural Development (733 MZ twins and 376 DZ twins) as if they were a sample of singletons, but allowing for their within-pair correlation by the use of cluster robust standard errors, the estimated effect of time 1 life-events on behaviour score at time 2, adjusting for behaviour score at time 1, is to increase the later behavioural score by 0.159 SD (95% CI 0.063, 0.253). This estimate is potentially biased by uncontrolled confounding, both those shared and non-shared by twins. Applying the standard BW model with the adjustment for prior behaviour for only the target twin gave a much smaller estimate (0.069; 95% CI: -0.068,0.324) which while controlling for shared confounders potentially induces collider bias as demonstrated by Sjolander et al (2012). Adding the prior behaviour score of the cotwin as an additional confounder removes this bias, and provides an estimate of the effect per life-event of 0.134SD (95%CI: 0.006, 0.262).Variations of the model of Figure 8 can be applied to contemporaneously measured phenotypes over time to assess direction of causality and extend the model to deal with mediation pathways while allowing control of genetic and shared environmental confounders for each of the exposure-mediator, mediator-outcome and the conditional exposure-outcome paths. Further extension to consider cross-lagged models is possible (Rommel et al. 2015). The use of IVs and data on related individuals can be combined, which in the case of Mendelian randomisation allows for potential adjustment for pleiotropic effects of the IV on exposure and outcome. The genotyping of both parents and children allows a powerful combination of the measured genes and family approaches in that one can distinguish transmitted and non-transmitted alleles from parents, the latter not contributing to the child's genetic risk but only to environmental exposure (Cheesman et al. 2020). Care is however called upon in order to avoid reverse causality from parents to offspring (Burgess et al, 2021) and, as for all IV based approaches, a full explication of the implicit assumptions (Swanson and Hernan 2013).

# 5. REPLICATION AND GENERALIZATION

## 5.1. Replication

Even a rigorous causal analysis of a high-quality study of substantial size is rarely sufficient by itself as proof of a causal effect nor as a valid estimate of the causal effect in another setting. Some form of multiplicity of evidence is necessary. The Reproducibility Project has highlighted how fragile are many, some long-held, beliefs of causation, and we are now seeing publication of many more replication studies of the kind that would previously never have seen the light-of-day. Rarely do these replications involve an exact replication of the sample setting and design, measurement battery, measurement schedule, implementation protocol, response rates and data analysis. Differences in the measures used can present challenges (Open Science Collaboration 2015). For example, when pooling and testing for heterogeneity of estimates using an effect-size scale may not correspond to using a common causal effect scale (Mathur and VanderWeele 2019). Some authors suggest that exact replication is not evidence of reproducibility (Drummond 2009), being unable to provide the evidence for the robustness of the relationship that should be expected, and that what we often need is evidence for the proof of the causal effect of change in the construct rather than change in the specific measure. A recognition that no study is perfect, that each and every one has some methodological flaw, argues for the need for triangulation, the bringing together of several studies, each with different, but hopefully unrelated, sources of bias (Lawlor, Tilling, and Davey Smith 2016). They propose that the different approaches address the same underlying causal question; for each approach the duration and timing of exposure that it assesses is taken into account when comparing results; that the key sources of bias are explicitly acknowledged when comparing results and that for each the expected direction of all key sources of potential bias are made explicit where this is feasible, and ideally within the set of approaches being compared there are approaches with potential biases that are in opposite directions. These conditions make it clear that this perspective remains some distance from the more informal seeking of support for a causal effect by triangulation of evidence from methodologically quite distant studies, such as human and animal studies.

## 5.2. Generalization

The considerations of the previous paragraph do not directly address the question of what the causal effect would be within a different setting. The focus is more strongly, sometimes exclusively, on internal validity, with much less consideration as to external validity. Stuart, in various articles has highlighted the importance of the latter, and that consideration of both is required, proposing the term target validity as a combining of the two (Westreich et al. 2019). One aspect influencing transportability of estimates relates to the fact that estimates of an average effect in one population may vary if the distribution of effect is different within the new target population. Ackerman et al (2021) propose the use of methodology from complex survey research to reweight estimates on stratification factors thought to be relevant to this variation (Ackerman et al. 2021). Where concerns remain in relation to potential residual confounding their distribution within the new target population may remain relevant, for example for inclusion in a propensity score adjustment (Dugoff, Schuler, and Stuart 2014).

# 6. CONCLUSIONS

While the distinction between association and cause has been a perennial concern in epidemiology, in the last two decades there has been an increasing shift away from the prevailing criteria (actually "guidelines") to assess causality outlined by Austin Bradford Hill in 1965, who presented them as "viewpoints from all of which we should study association before we cry causation" (Hill 1965). The shift is towards a more precise definition of what "effect" is being targeted. This has revealed unexpected multiplicity and complexity, especially in the setting of long-term causation typical of life course research. Choosing an appropriate design and estimator are important, but these should be preceded by choosing the right estimand which ideally should correspond to the comparisons at the heart of the substantive question being investigated.

Analysis in public health should be about informing us of the likely impact of alternative actions and policy decisions. It is therefore surprising that so much of what is presented specifies what these alternatives are with considerable imprecision. The counterfactual approach attempts to remove this imprecision, specifying exactly the hypothetical scenarios to be compared, and the choice of assumptions you may have for identifying and estimating their contrasts. At times, this approach can be seen as over-elaborate, but we would argue that its influence is having a variety of desirable effects. First, in encouraging an approach to analysis that focuses on targeting well-defined alternatives, this should both help clarify and specify policy alternatives and help make the relevance of the analysis to policy making more transparent. Second, the specification of the factual is as relevant as the counterfactual. This is especially true for pooled and meta-analyses where greater attention should be being paid that the estimates of effects that are being combined correspond to the same contrast of counterfactuals (Elango et al. 2015). Forcefully recognised by those attempting to reconcile causal effect estimates from epidemiological and RCT studies, this concern for consistency of the exposure definition is central to those attempting trial emulation from observational data. It has also been found that a number of estimators of causal effects proposed for observational and twin data, either make implausible assumptions that were laid bare by counterfactual expositions. The traditional focus on distinguishing among life course models has also been found not to be very informative for investigate alternative interventions, while understanding mechanisms by exploiting developments in mediation analysis may be more fruitful.

We have seen how counterfactuals can be derived from otherwise similar individuals but who experienced different levels of exposure, or the same individual but with periods of different exposure, or on siblings with similarities and differences in exposure and so on. Each derivation requires different assumptions to be satisfactorily interpretable as counterfactual, with sometimes the task having to be recognised as impossible. Some estimators of counterfactual-based contrasts can use information from only a subset of the available sample data – for example from only those that are "matchable" or not already at the maximum (or minimum) level of the policy manipulable variable or have experienced change in the focal exposure. Inferring that the causal estimate applies to those excluded is clearly a separate inferential step, so identifying the population to whom a causal estimate applies is thus also important, with generalisation to other populations requiring both careful thought as to the equivalence of the distributions and likely some additional analysis and adjustment methods such as weighting.

We have tried to show that the estimation of life course relevant counterfactual based estimands requires not just appropriate consideration of assumptions but the exploitation of various study designs that can allow estimators or generalization that require different assumptions. A feature of much causal inference has been to make as few parametric assumptions as possible, including minimizing assumptions as to functional forms. This may restrict the value of the work for prediction. Heckman and Pinto (2015) argue that the counterfactual approach lacks models of mechanism, that it focuses on effects of causes rather than causes of effects. Therefore, it can have nothing to say about the effects of previously unexperienced levels of risks, or entirely novel risks or policy initiatives, for example of the kind that the COVID-19 pandemic presented so plentifully. In the context where the counterfactual may be entirely hypothetical, such as involving the extrapolation of exposures into currently unobserved ranges, a purely theoretical justification of the necessary assumptions is required but can sometimes still be persuasive. This brings us to the final conclusion, that these endeavours have some unsurmountable challenges, especially when the scope is as broad ranging as the last one discussed. In each case, contextual knowledge is an unavoidable requirement, without which counterfactuals cannot be defined, nor sources of bias recognised. In presenting structures of thought, assumptions and methods that are of general relevance, considerable abstraction is inevitable. However, their meaningful application requires making full use of contextual knowledge, without which useful counterfactuals cannot be defined, nor the implications of assumptions understood, nor potential sources of bias recognised.

# Acknowledgments

We are extremely grateful to all the families who took part in the studies, the midwives for their help in recruiting them, and the whole ALSPAC team, which includes interviewers, computer and laboratory technicians, clerical workers, research scientists, volunteers, managers, receptionists and nurses. Thanks also to the VTSABD families and study team.

Funding

AP was partially supported by the NIHR (NF-SI-0617-10120) and the Biomedical Research Centre at South London and Maudsley NHS Foundation Trust and King's College London. The views expressed are those of the authors and not necessarily those of the UK NHS, NIHR or the Department of Health and Social Care. MH is funded by a fellowship from the Medical Research Council UK (MR/T027843/1). BDS is partially supported by the MRC Methodology Grant: MR/R025215/1.

The UK Medical Research Council and Wellcome (Grant ref: 217065/Z/19/Z) and the University of Bristol provide core support for ALSPAC. This publication is the work of the authors and BDS and MH will serve as guarantors for the contents of this paper. A comprehensive list of grants funding is available on the ALSPAC website (http://www.bristol.ac.uk/alspac/external/documents/grant-acknowledgements.pdf.

# TABLES

**Table 1.** Overview of conceptual life course models and their corresponding causal contrasts expressed in terms of potential outcomes, for a setting where exposure is measured at two time points.

| Life course Model | Specification | Relevant estimands | Comment |
|---|---|---|---|
| Cumulative exposure | The exposure causally affects the outcome at both time points with similar magnitude | $CDE_1(a_2)$ $TCE_2$ | If model is correct, the estimands are equal |
| Sensitive period | The exposure causally affects the outcome at both time points but with different magnitude | $CDE_1(a_2)$ $TCE_2$ | If model is correct, one of the estimands is larger than the other |
| Critical period | The exposure causally affects the outcome at only one time point t | $CDE_1(a_2)$ $TCE_2$ | If model is correct, one of the estimands is equal to zero |
| Pathway | The earlier exposure causally affects the outcome but with an intensity that depends on the later exposure | $CDE_1(a_2)$ | If model is correct, the estimand varies with values taken by $A_2$ |

Abbreviations: CDE: Controlled direct effect, TCE: Total causal effect; suffices indicate the timing of exposure

**Table 2.** Estimates and 95% confidence intervals (CI) for the total causal effect (*TCE*) of standardized birth weight and BMI at 12 years, and the controlled direct effect (*CDE*) of standardized birth weight for different values of BMI at 12 years[†]; ALSPAC Study, N=1,953

| Exposure | Estimand | Controlling for baseline confounders Estimate (SE) | Controlling for baseline and intermediate confounders Estimate (SE) |
|---|---|---|---|
| BMI at 12 years (standardized) | $TCE_2$ | 0.244 (0.018) | — |
| Birth weight (standardized) | $CDE_1(0)$ | 0.020 (0.019) | 0.023 (0.016) |
|  | $CDE_1(1)$ | 0.027 (0.018) | 0.031 (0.017) |
|  | $CDE_1(-1)$ | 0.013 (0.021) | 0.017 (0.020) |

[†]Estimation of $CDE_1(a_2)$ was performed by g-computation controlling for baseline confounders; $TCE_2$ was additionally controlled for birth weight. Standard errors were derived from 1000 bootstraps samples.

Abbreviations: *TCE*: Total causal effect, *CDE*: Controlled direct effect; CI: confidence interval; BMI: body mass index, ALSPAC: Avon longitudinal studies of parents and children
Baseline confounders: maternal education, family occupation, maternal smoking during pregnancy; maternal pre-pregnancy BMI, maternal pre-pregnancy history of psychopathology.
Intermediate confounder: BMI at age 7 years.

**Table 3.** Estimates and 95% confidence intervals (CI) for the total causal effect (*TCE*) of standardized birth weight and the interventional direct and indirect effect of birth weight via BMI at 12 years; ALSPAC Study, N=1,953.

| Estimands | Allowing for non-linear relationships | Assuming linear relationships |
|---|---|---|
| | Estimate (SE) | Estimate (SE) |
| *TCE* | 0.047 (0.016) | 0.030 (0.012) |
| *IDE* | 0.023 (0.016) | 0.008 (0.011) |
| *IIE* | 0.024 (0.007) | 0.023 (0.006) |

†Estimation was performed by g-computation controlling for baseline and intermediate confounders. Standard errors derived from 1000 bootstraps samples.

Abbreviations: *TCE:* Total causal effect, *IDE:* Interventional direct effect; *IIE:* Interventional indirect effect; CI: confidence interval; BMI: body mass index

Baseline confounders: maternal education, family occupation, maternal smoking during pregnancy; maternal pre-pregnancy BMI, maternal pre-pregnancy history of psychopathology.

Intermediate confounder: BMI at age 7 years.

**Table 4.** Estimates and 95% confidence intervals (CI) for the total causal effect (*TCE*) of standardized birth weight and the interventional direct and indirect effect of birth weight via BMI growth from age 7 to 12 years; ALSPAC Study, N=1,953.

| Estimands | Allowing for non-linear relationships | Assuming linear relationships |
|---|---|---|
| | Estimate[†] (SE) | Estimate[††] (SE) |
| TCE | 0.045 (0.012) | 0.031 (0.011) |
| IDE | 0.018 (0.011) | 0.004 (0.011) |
| IIE -all | 0.027 (0.005) | 0.027 (0.004) |
| $\quad$ IIE$_1$-Via BMI from 7 to 9y and upstream | -0.002 (0.004) | -0.001 (0.002) |
| $\quad$ IIE$_2$-Via BMI from 10 to 12y and upstream | 0.029 (0.006) | 0.028 (0.011) |
| $\quad$ Remainder | -0.00002 ($10^{-6}$) | – |
| TCE | 0.045 (0.011) | 0.031 (0.012) |
| IDE | 0.017 (0.011) | 0.006 (0.011) |
| IIE -all | 0.028 (0.005) | 0.024 (0.006) |
| $\quad$ IIE$_1$-Via latent size and upstream | 0.029 (0.004) | 0.026 (0.005) |
| $\quad$ IIE$_2$-Via latent velocity and upstream | -0.0004 (0.001) | -0.002 (0.002) |
| $\quad$ Remainder | -0.00002 ($10^{-6}$) | – |

[†]Estimation was performed by g-computation controlling for confounders. Standard errors derived from 1,000 bootstraps samples.

[††] Estimation was performed by maximum likelihood controlling for confounders. Standard errors derived by the delta method.

Abbreviations: *TCE:* Total causal effect, *IDE:* Interventional direct effect; *IIE:* Interventional indirect effect; CI: confidence interval; BMI: body mass index

Baseline confounders: maternal education, family occupation, maternal smoking during pregnancy; maternal pre-pregnancy BMI, maternal pre-pregnancy history of psychopathology.

**FIGURES**

Legends

**Figure 1.** Conceptual models corresponding to the two explanations for the association between childhood SES ($A_1$) and later life cognition ($Y$): (a) Childhood SES as precursor of the true cause, adult SES ($A_2$); (b) Childhood SES as the cause; (c) both as causes.

**Figure 2.** (a) Causal DAG of the association between childhood SES ($A_1$), adult SES ($A_2$), and later life cognition ($Y$), with confounders $C$ and $L$, and the binary indicator $R$ capturing whether study members are observed ($R=1$) or missing ($R=0$).The square around $R$ indicates that analysis are restricted to complete records. (b) Causal directed acyclic graph of the association between childhood SES ($A_1$), adult SES ($A_2$), and later life cognition ($Y$), with $A^*_2$ being unobserved (indicated by a circle) but proxied by the variable $A_2$.

**Figure 3.** (a) Causal DAG of the association between childhood birth weight (BW), BMI at 12y ($BMI_{12}$), and binge eating score at 13.5y (*BE*), with *C* representing confounders. (b) Expanded DAG that includes an intermediate confounder *L*.

**Figure 4.** Causal DAG of the association between childhood birth weight (BW), childhood BMI, and binge eating score at 13.5y (*BE*), with *C* representing baseline confounders and L intermediate confounders. Not all arrows are included to aid the visual display.

**Figure 5.** Causal DAG of the association between an exposure X, a time-varying mediator represented by a latent intercept ($I_M$) and a latent slope ($S_M$), and a latent outcome represented by a latent intercept ($I_Y$) and a latent slope ($S_Y$). Double headed arrow represents correlation indued by a common factor (not included for simplicity).

**Figure 6.** Causal DAG of the association between childhood birth weight (BW), latent intercept and slope of the childhood BMI measures, and binge eating score at 13.5y (BE), with C representing baseline confounders and L intermediate confounders.

**Figure 7.** (a) Causal DAG depicting the setting with a randomised encouragement to uptake an intervention, the intervention *A*, the outcome Y, unmeasured confounders U and measured confounders C ; (b) Causal DAG depicting the setting with a randomised encouragement to uptake an intervention, the intervention *A*, a mediator M, outcome Y and measured and unmeasured confounders U and C, with the addition of an indicator of the interaction between encouragement and measured confounders C; (c) Causal DAG depicting the setting with an instrumental variable Z, an exposure A measured at two time points ($A_0$, $A_1$), an outcome *Y* and unmeasured confounders *U*; (d) Causal DAG depicting the setting with an instrumental variable Z, an exposure A measured at two time points ($A_0$, $A_1$), an outcome *Y* measured at two time points ($Y_0$, $Y_1$), and unmeasured confounders *U*.

**Figure 8.** Causal DAG of the relationship between life events X and behaviour Y at two time points for two twins: $X_1$ life event for twin 1, $X_2$ life event for twin 2, $Y_{11}$ and $Y_{12}$ behaviour for twin1 at time points 1 and 2; $Y_{21}$ and $Y_{22}$ for twin 2, U shared unmeasured confounders.

**Figure 1.** Conceptual models corresponding to the two explanations for the association between childhood SES ($A_1$) and later life cognition ($Y$): (a) Childhood SES as precursor of the true cause, adult SES ($A_2$); (b) Childhood SES as the cause; (c) both as causes.

(a)

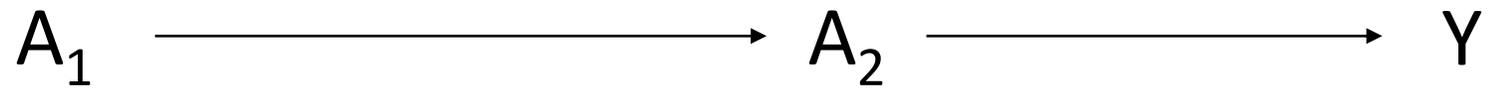

(b)

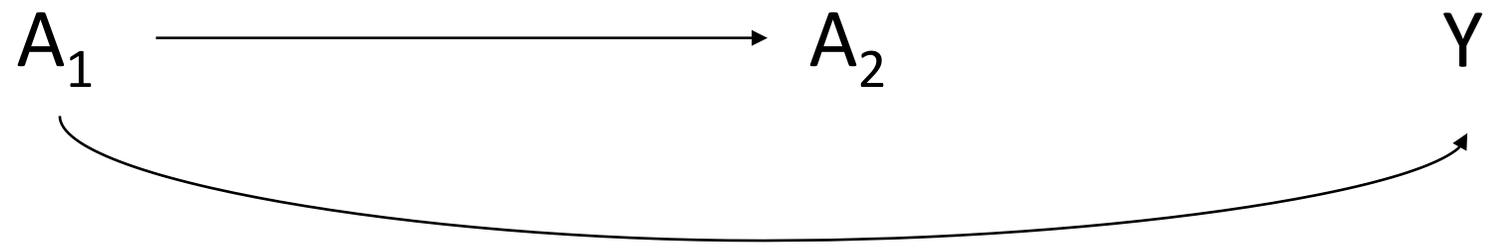

(c)

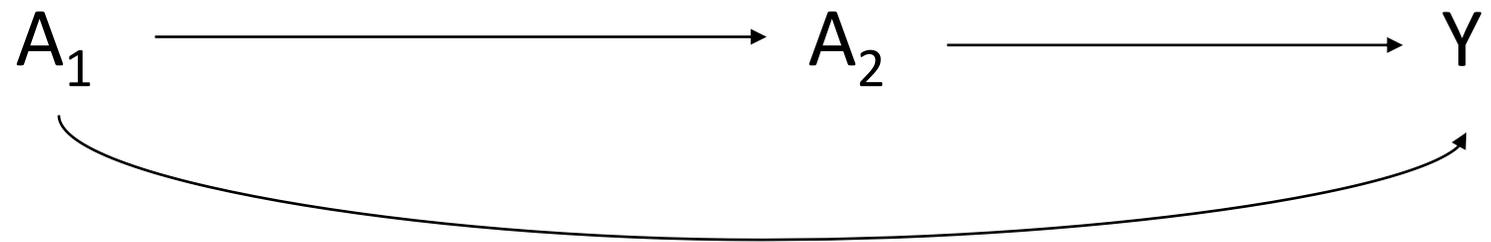

**Figure 2.** (a) Causal DAG of the association between childhood SES ($A_1$), adult SES ($A_2$), and later life cognition ($Y$), with confounders $C$ and $L$, and the binary indicator $R$ capturing whether study members are observed ($R=1$) or missing ($R=0$). The square around $R$ indicates that analysis are restricted to complete records. (b) Causal directed acyclic graph of the association between childhood SES ($A_1$), adult SES ($A_2$), and later life cognition ($Y$), with $A^*_2$ being unobserved (indicated by a circle) but proxied by the variable $A_2$.

(a)

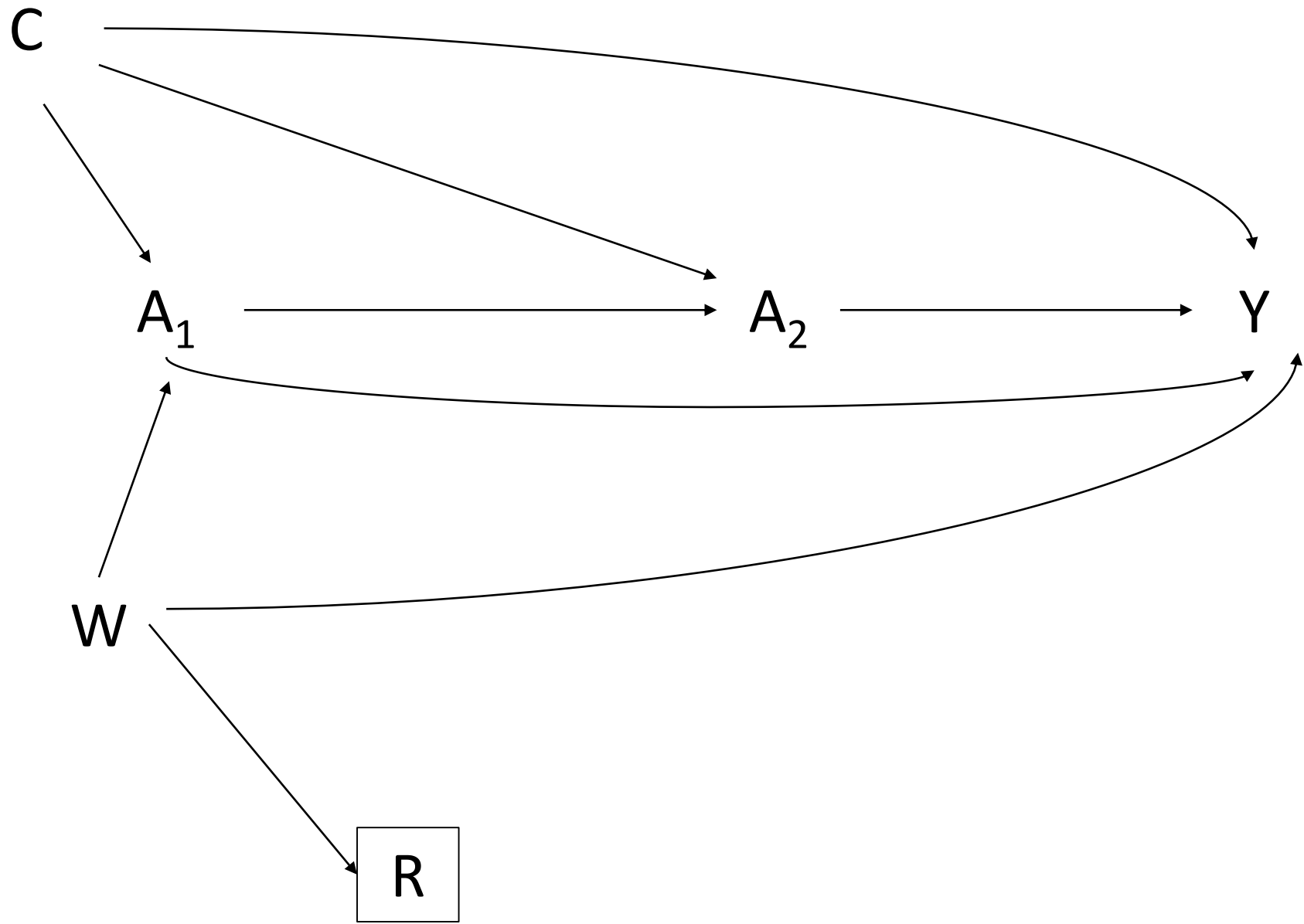

(b)

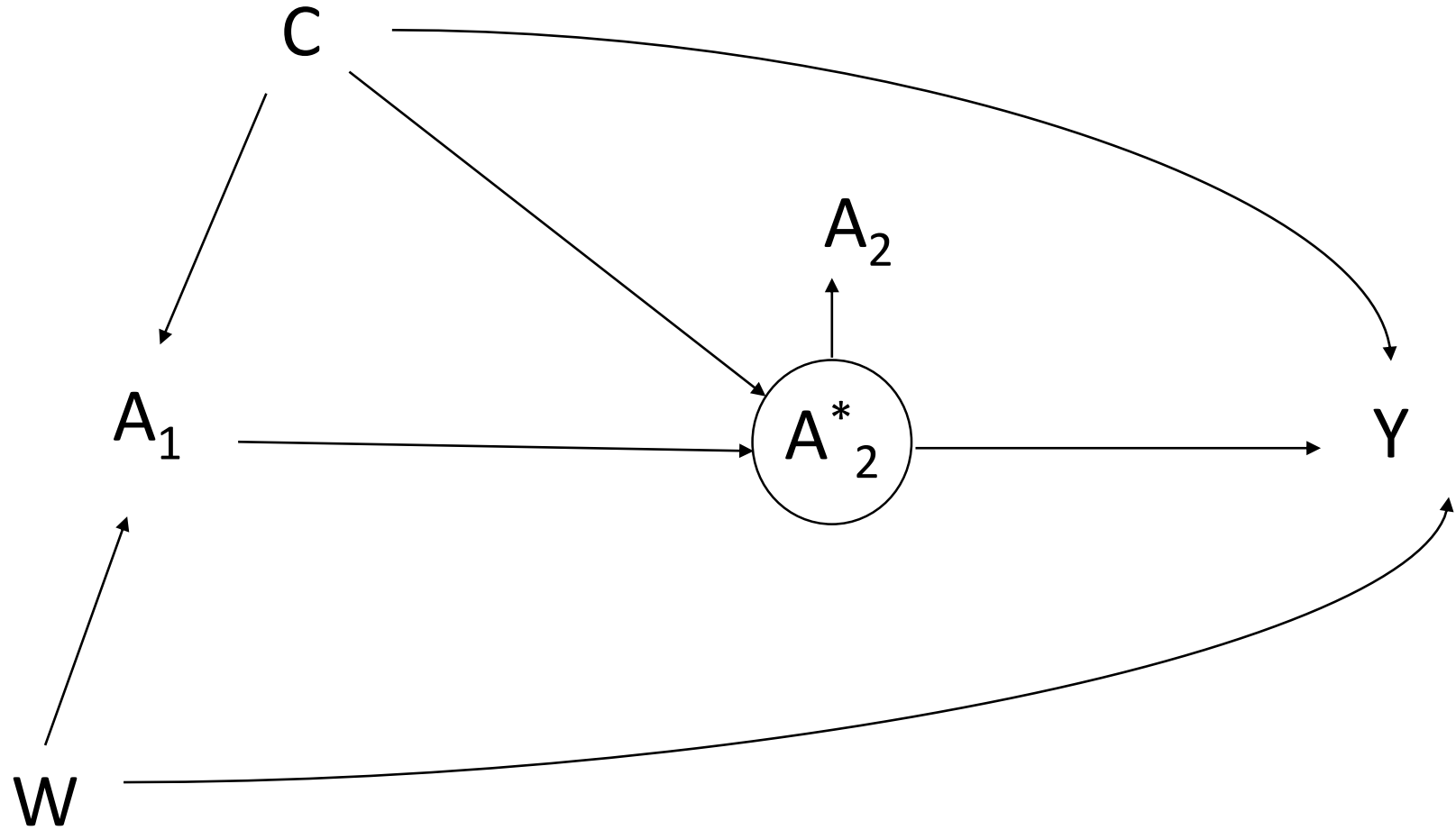

**Figure 3.** (a) Causal DAG of the association between childhood birth weight (BW), BMI at 12y (BMI$_{12}$), and binge eating score at 13.5y (BE), with C representing confounders. (b) Expanded DAG that includes an intermediate confounder L.

(a)

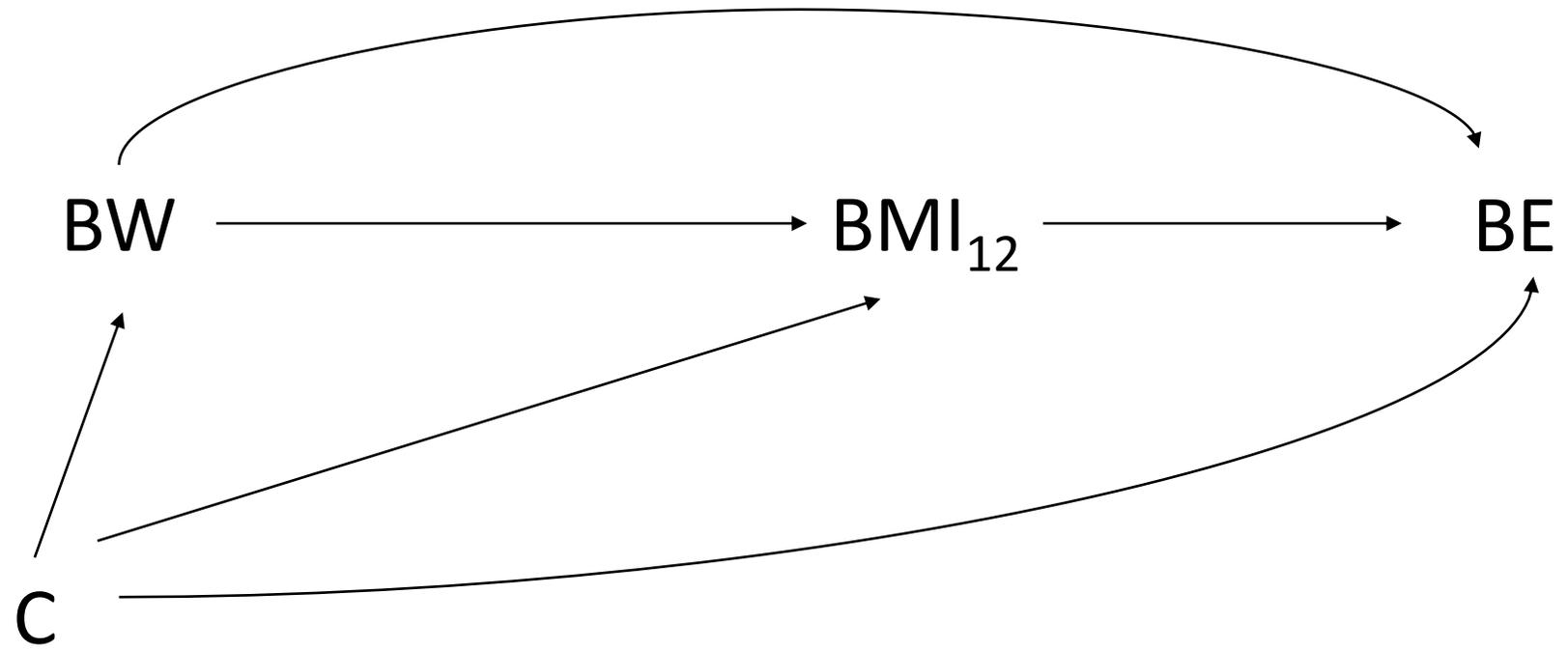

(b)

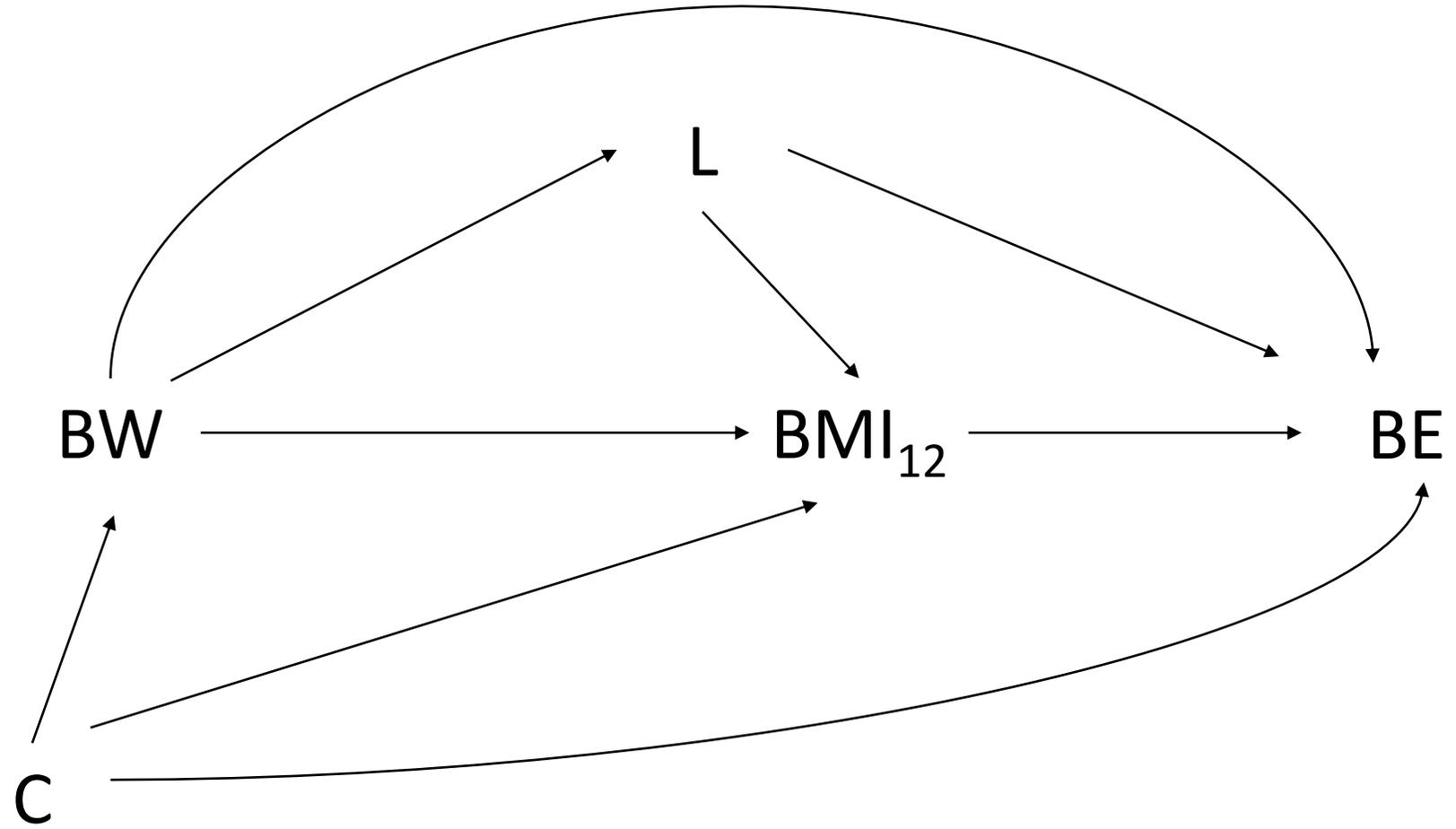

**Figure 4.** Causal DAG of the association between childhood birth weight (BW), childhood BMI, and binge eating score at 13.5y (*BE*), with *C* representing baseline confounders and L intermediate confounders. Not all arrows are included to aid visual clarity.

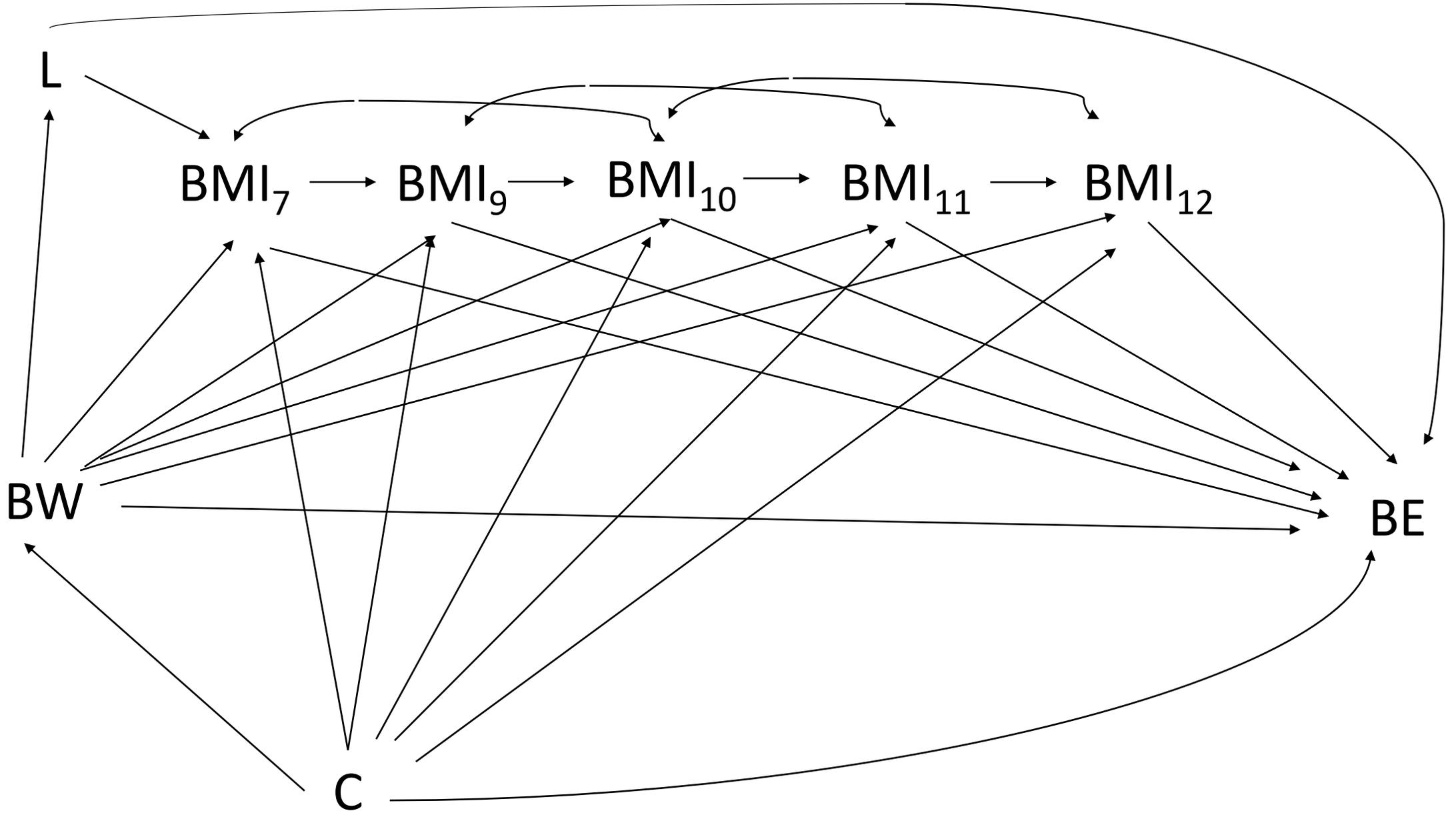

**Figure 5.** Causal DAG of the association between an exposure X, a time-varying mediator represented by a latent intercept ($I_M$) and a latent slope ($S_M$), and a latent outcome represented by a latent intercept ($I_Y$) and a latent slope ($S_Y$). Double headed arrow represents correlation indued by a common factor (not included for simplicity).

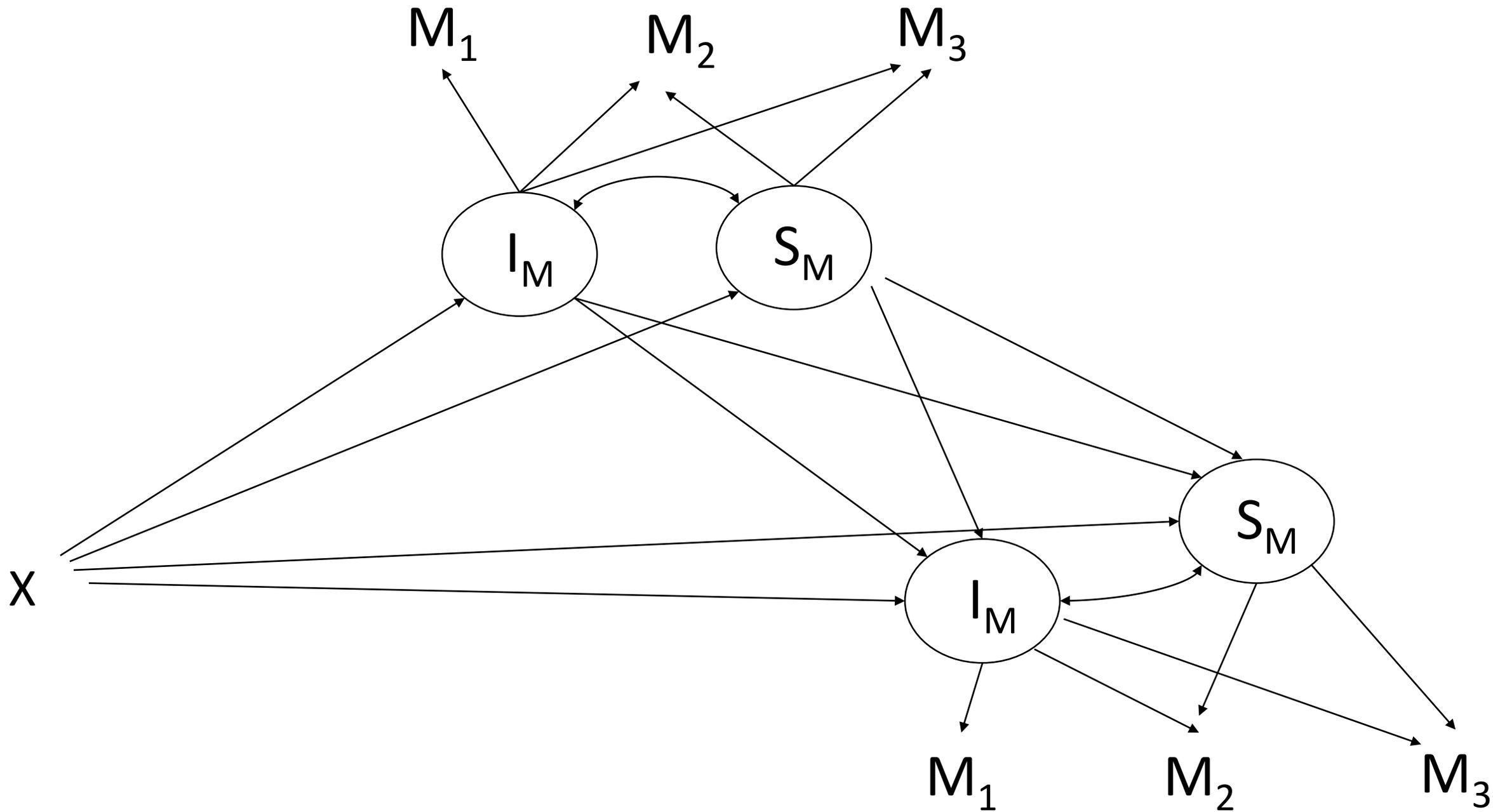

**Figure 6.** Causal DAG of the association between childhood birth weight (BW), latent intercept and slope of the childhood BMI measures, and binge eating score at 13.5y (BE), with C representing baseline confounders and L intermediate confounders.

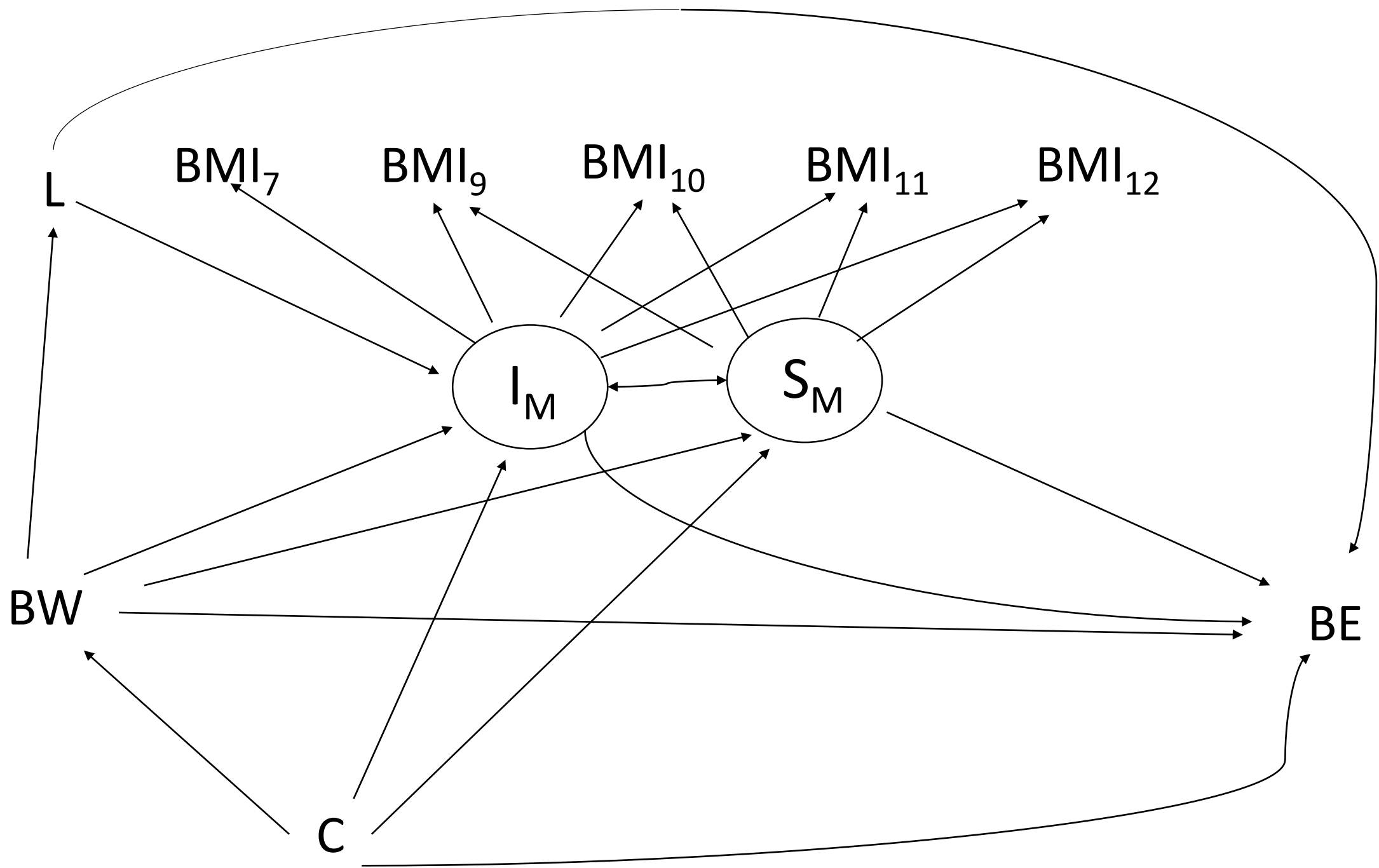

**Figure 7.** (a) Causal DAG depicting the setting with a randomised encouragement to uptake an intervention, the intervention A, the outcome Y, unmeasured confounders U and measured confounders C ; (b) Causal DAG depicting the setting with a randomised encouragement to uptake an intervention, the intervention A, a mediator M, outcome Y and measured and unmeasured confounders U and C, with the addition of an indicator of the interaction between encouragement and measured confounders C; (c) Causal DAG depicting the setting with an instrumental variable Z, an exposure A measured at two time points ($A_0$, $A_1$), an outcome Y and unmeasured confounders U; (d) Causal DAG depicting the setting with an instrumental variable Z, an exposure A measured at two time points ($A_0$, $A_1$), an outcome Y measured at two time points ($Y_0$, $Y_1$), and unmeasured confounders U.

(a)

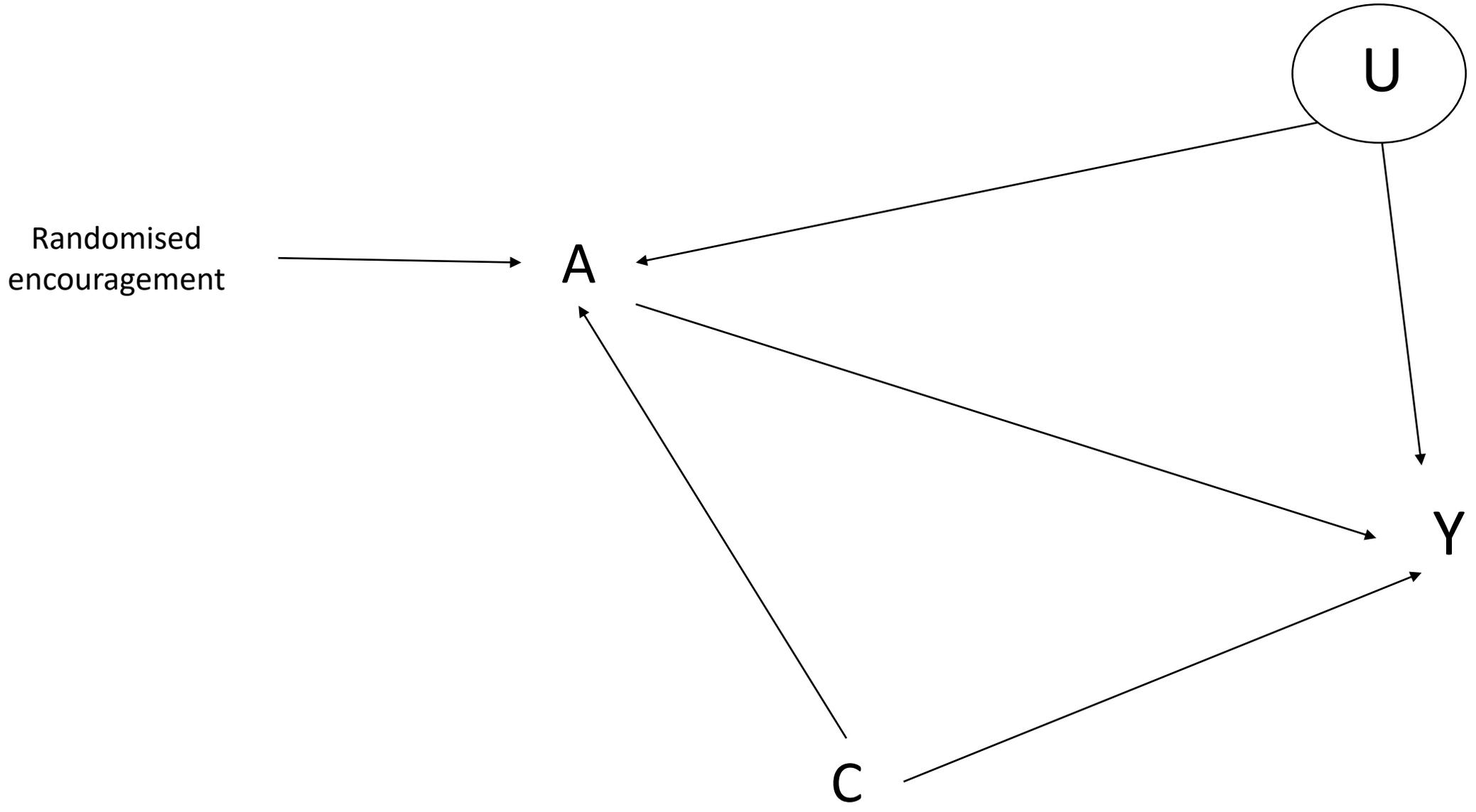

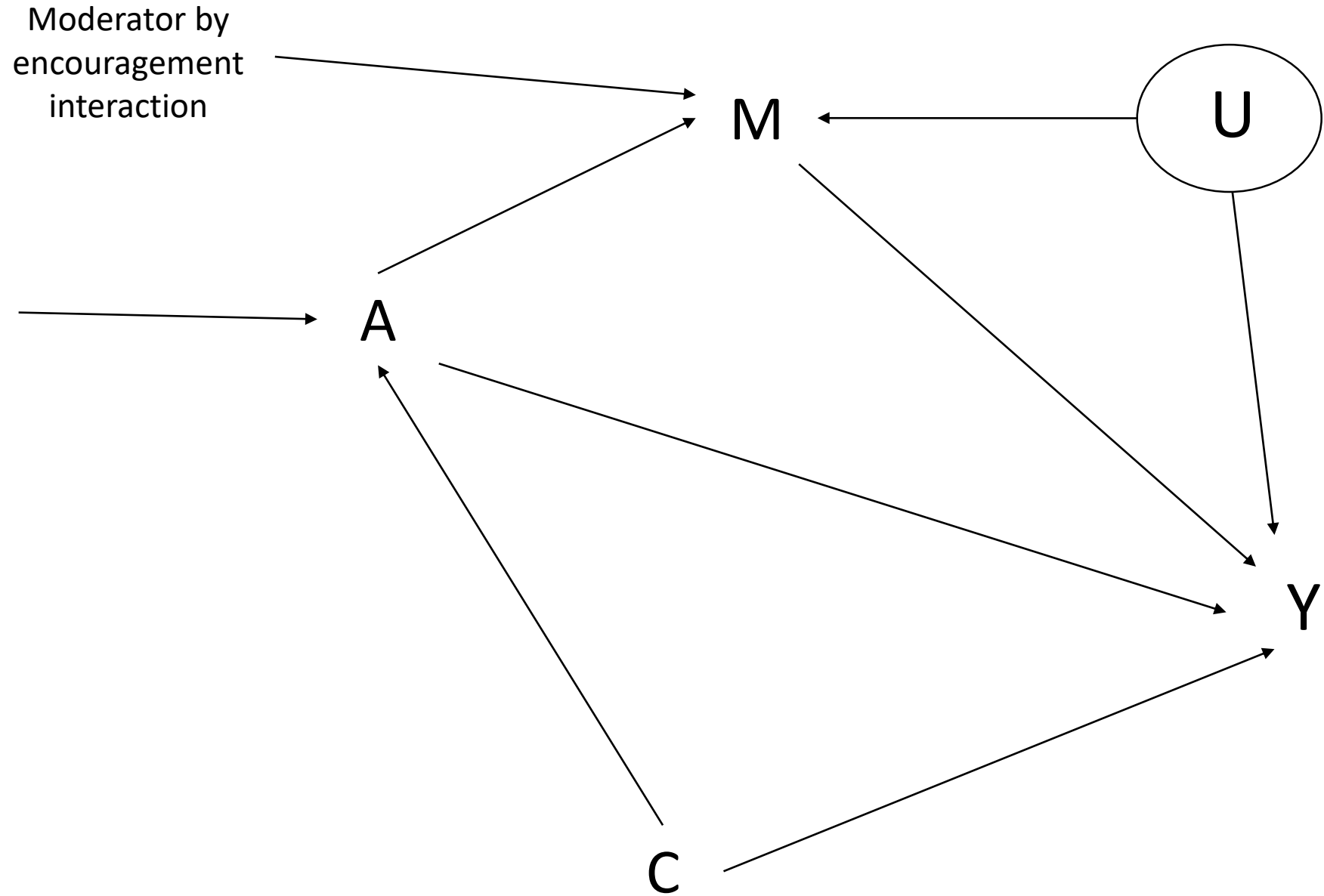

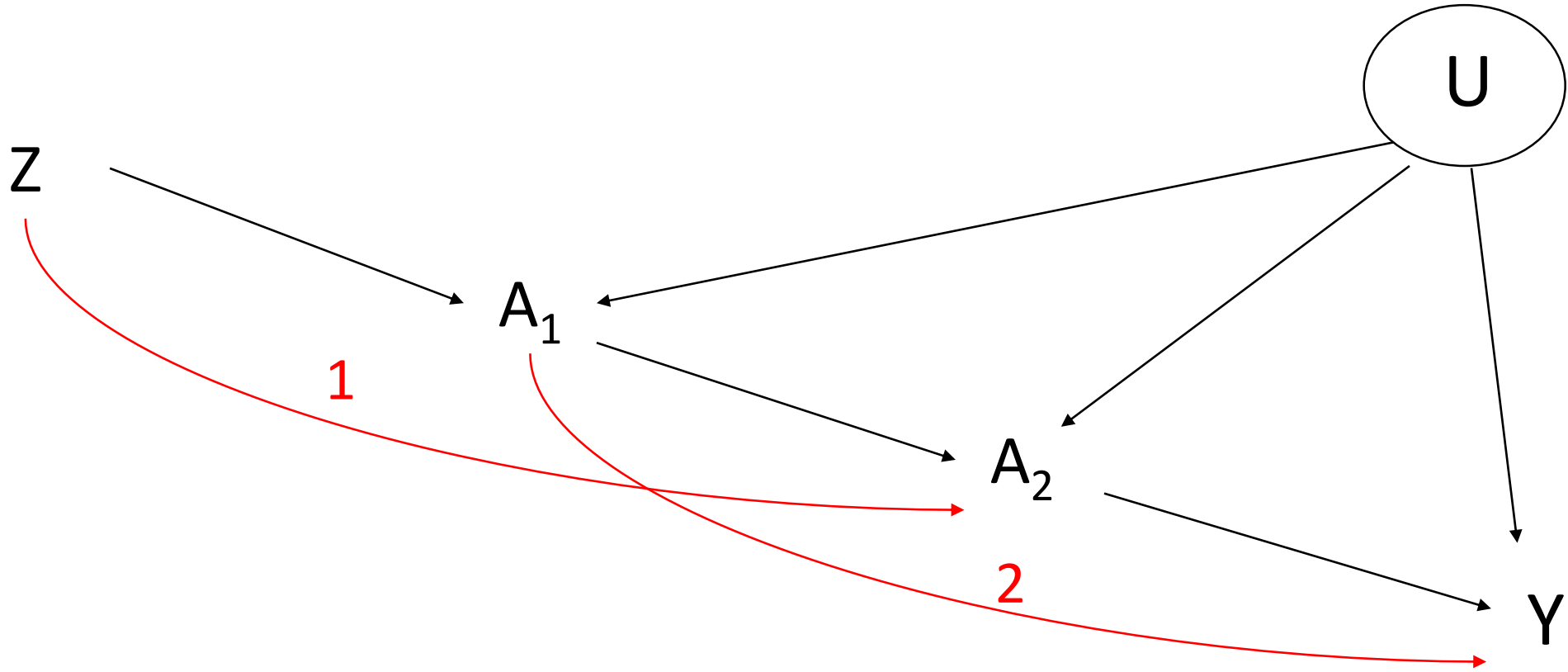

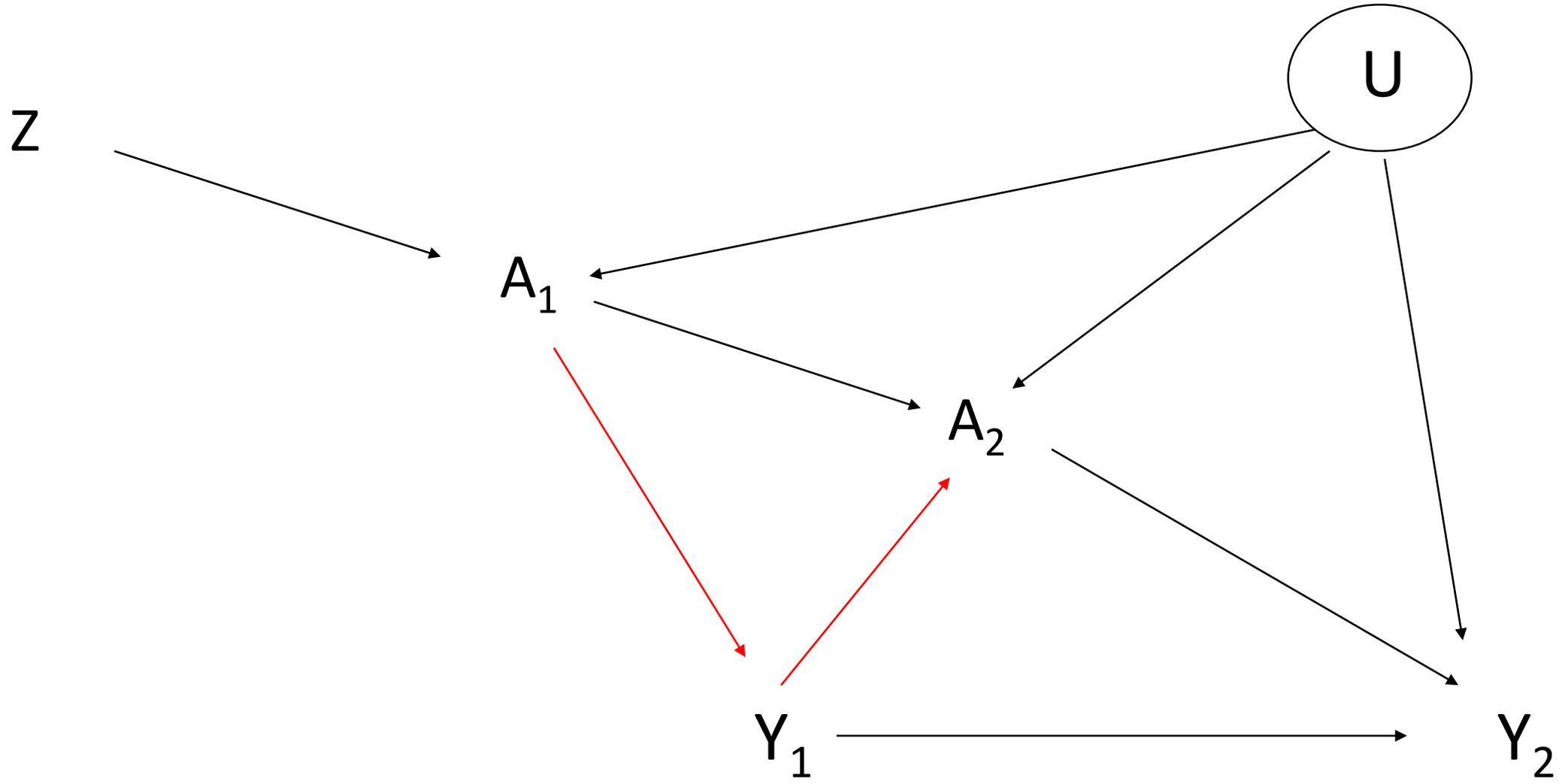

**Figure 8.** Causal DAG of the relationship between life events X and behaviour Y at two time points for two twins: $X_1$ life event for twin 1, $X_2$ life event for twin 2, $Y_{11}$ and $Y_{12}$ behaviour for twin1 at time points 1 and 2; $Y_{21}$ and $Y_{22}$ for twin 2, U shared unmeasured confounders.

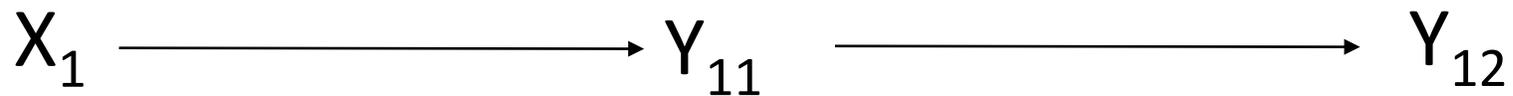
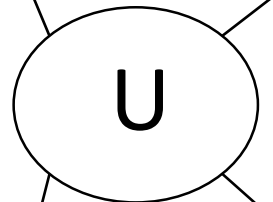
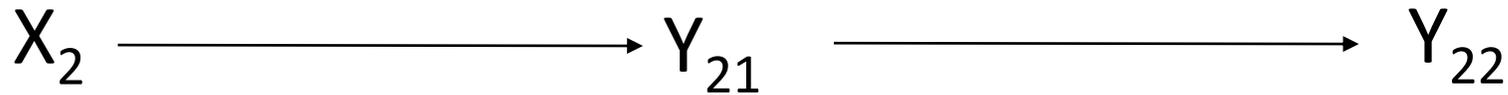

# APPENDIX

## A.1 Data

**Avon Longitudinal Study of Parents and Children (ALSPAC)**

Participants included in this study are a subsample of adolescents of the population-based ALSPAC cohort that recruited pregnant women in the southwest of England (Boyd, Golding et al. 2013, Fraser, Macdonald-Wallis et al. 2013). All pregnant women that were expected to have a child in the period of 1 April 1991 until 31 December 1992 were contacted to participate in the original cohort. At the beginning, 14,451 pregnant women took part, and 13,988 children were alive at the end of year one. To guarantee independence of individuals, one sibling per set of multiple births (n = 203 sets) is randomly included in our sample. Please note that the study website contains details of all the data that are available through a fully searchable data dictionary and variable search tool and reference the following webpage: http://www.bristol.ac.uk/alspac/researchers/our-data/.

Ethical approval for the ALSPAC participants was obtained from the ALSPAC Ethics and Law Committee and the Local Research Ethics Committees: www.bristol.ac.uk/alspac/researchers/research-ethics/. Informed consent for the use of data collected via questionnaires and clinics was obtained from participants following the recommendations of the ALSPAC Ethics and Law Committee at the time.

MEASURES

<u>First Exposure:</u> birth weight (grams) was obtained from obstetric records.

<u>Later exposure/Mediators</u>: Body mass index (BMI; in kg/m2), objectively measured up to six times when participants were (around) 7.5, 8.6, 9.8, 10.6, 11.8 (referred to as 7,9,10,11,12) years. Height was measured to the nearest millimetre with the use of a Harpenden Stadiometer (Holtain Ltd.). Weight was measured with a Tanita Body Fat Analyzer (Tanita TBF UK Ltd.) to the nearest 50 g.

<u>Outcome:</u> Parentally-reported ED behaviors (p-ED: at mean child age 13.1 years (standard deviation, SD=0.2), data were collected via the Developmental and Well-being assessment (DAWBA), a semi-structured validated interview that generates a range of psychiatric diagnoses in children and adolescents (Goodman, Ford et al. 2000). The ED section of the DAWBA was given to parents and comprises 28 questions on ED behaviors and cognitions. These were used to derive three disordered eating patterns: 1. Binge eating/overeating; 2. shape and weight concern and weight control behaviors, and 3. food restriction, using exploratory structural equation modeling. Data on these three patterns were available on 3,529 girls. Because these are latent factors derived from structured questionnaires, they are standardized measures with mean 0 and standard deviation (SD) of 1. Only the binge eating score is used in this paper.

<u>Confounders</u>: High maternal education at birth of child was defined by mothers having completed education up to A-Levels, the requirement for applying to university in the UK. Maternal age and lowest parental social class were obtained at enrolment. At 12 weeks gestation, women were asked about any recent or past history of severe depression, schizophrenia, alcoholism, anorexia nervosa, bulimia nervosa, and other psychiatric disorders. Multiple answers were possible; therefore, women could report more than one disorder. This information was combined into a variable indicating presence of any pre-pregnancy psychopathology.

Analyses were restricted to 1,953 girls with complete birth weight, binge eating scores, confounders and at least one BMI measure.

**Virginia Twin Study of Adolescent Behavioural Development (VTSABD)**

SAMPLE: The VTSABD is an ongoing cohort-longitudinal study of twins born between 1974 and 1983. The study was limited to white families because there were too few data from other ethnic groups to provide adequate power to detect differential effects. A sample of 6,837 putative twin pairs born between 1970 and 1985 was ascertained through the state school system and participating private schools in Virginia, and through families who contacted the VTSABD. Parents of these twins were sent a brief letter describing the study and asked to complete a questionnaire regarding zygosity. Non-responders were followed up with 2 further mailings and attempted telephone contact. After the contact, 1,424 putative pairs were removed for a variety of reasons, including having moved out of state, having moved and left no forwarding address, the children not being twins, and duplicate ascertainment. Of the 5,413 twin families, 1,894 were selected for interview in the main study. Others were excluded because they were not white, were outside the age range, had moved out of state by the time of interviewing, or had moved without leaving a forwarding address. A total of 1,412 families (2824 children) participated in the first wave, 75% of the targeted sample 2 subsequent waves of data collection occurring at approximately 1 year intervals (Meyer, Silberg et al. 1996).

MEASURES: The principal measures used in our analysis were the child reported total life-events score from the Life-Events Questionnaire (Johnson & McCutcheon, 1980; Johnson, 1986) and the total behaviour score from the Olweus Aggression Scale (Olweus, 1989).

## A.2 Codes

Stata and Mplus codes used to produce the results reported in the examples are to be found here: https://github.com/bldestavola/ARSIA-lifecourse